\documentclass[notitlepage,twocolumn,letterpaper,natbib,aps,prd,amsmath,amsfonts,nofootinbib,preprintnumbers,superscriptaddress,secnumarabic]{revtex4-1}
\pdfoutput=1
\usepackage{amssymb,amsmath,latexsym,mathrsfs}
\usepackage{url}
\usepackage{enumitem}
\usepackage{graphicx}
\usepackage{booktabs}
\usepackage[usenames,dvipsnames]{color}
\usepackage[breaklinks,colorlinks,urlcolor=magenta,citecolor=magenta,linkcolor=magenta]{hyperref}
\usepackage{multirow}
\usepackage{float}
\usepackage{cases}
\usepackage{blindtext}
\usepackage{pifont}
\usepackage{hhline}
\usepackage{xcolor}

\definecolor{linkcolor}{rgb}{0.0, 0.47, 0.75}
\definecolor{citecolor}{rgb}{1.0, 0.5, 0.0}
\definecolor{linkcolor}{rgb}{0.390625,0.5607843137,0.99609375}
\hypersetup{
  linkcolor  = linkcolor,
  citecolor  = linkcolor,
  urlcolor   = linkcolor,
  colorlinks = true
}

\begin{document}

\title{Albatross: A scalable simulation-based inference pipeline \\ for analysing stellar streams in the Milky Way}

\author{James Alvey}
\email{j.b.g.alvey@uva.nl}
\thanks{ORCID: \href{https://orcid.org/0000-0003-2020-0803}{0000-0003-2020-0803}}
\affiliation{GRAPPA Institute, Institute for Theoretical Physics Amsterdam,\\
University of Amsterdam, Science Park 904, 1098 XH Amsterdam, The Netherlands}

\author{Mathis Gerdes}
\email{m.gerdes@uva.nl}
\thanks{ORCID: \href{https://orcid.org/0000-0002-2369-1436}{0000-0002-2369-1436}}
\affiliation{GRAPPA Institute, Institute for Theoretical Physics Amsterdam,\\
University of Amsterdam, Science Park 904, 1098 XH Amsterdam, The Netherlands}

\author{Christoph Weniger}
\email{c.weniger@uva.nl}
\thanks{ORCID: \href{https://orcid.org/0000-0001-7579-8684}{0000-0001-7579-8684}}
\affiliation{GRAPPA Institute, Institute for Theoretical Physics Amsterdam,\\
University of Amsterdam, Science Park 904, 1098 XH Amsterdam, The Netherlands}

\begin{abstract}

\noindent Stellar streams are potentially a very sensitive observational probe of galactic astrophysics, as well as the dark matter population in the Milky Way. On the other hand, performing a detailed, high-fidelity statistical analysis of these objects is challenging for a number of key reasons. Firstly, the modelling of streams across their (potentially billions of years old) dynamical age is complex and computationally costly. Secondly, their detection and classification in large surveys such as \emph{Gaia} renders a robust statistical description regarding \emph{e.g.}, the stellar membership probabilities, challenging. As a result, the majority of current analyses must resort to simplified models that use only subsets or summaries of the high quality data. In this work, we develop a new analysis framework that takes advantage of advances in simulation-based inference techniques to perform complete analysis on complex stream models. To facilitate this, we develop a new, modular dynamical modelling code \texttt{sstrax} for stellar streams that is highly accelerated using \texttt{jax}. We test our analysis pipeline on a mock observation that resembles the GD1 stream, and demonstrate that we can perform robust inference on all relevant parts of the stream model simultaneously. Finally, we present some outlook as to how this approach can be developed further to perform more complete and accurate statistical analyses of current and future data.

\vspace*{5pt} \noindent \textbf{\texttt{GitHub}}: Our new \texttt{jax}-accelerated stellar streams modelling code \texttt{sstrax} can be found \href{https://github.com/undark-lab/sstrax}{here}. The \texttt{swyft}-based inference library \texttt{albatross} is available at this \href{https://github.com/undark-lab/albatross}{link}.
\end{abstract}

\maketitle
\hypersetup{
  linkcolor  = linkcolor,
  citecolor  = linkcolor,
  urlcolor   = linkcolor
}

\section{Introduction}\label{sec:intro}

\begin{figure*}[t]
    \centering
    \includegraphics[width=\linewidth,trim={0.1cm 0.1cm 0.1cm 0.1cm},clip]{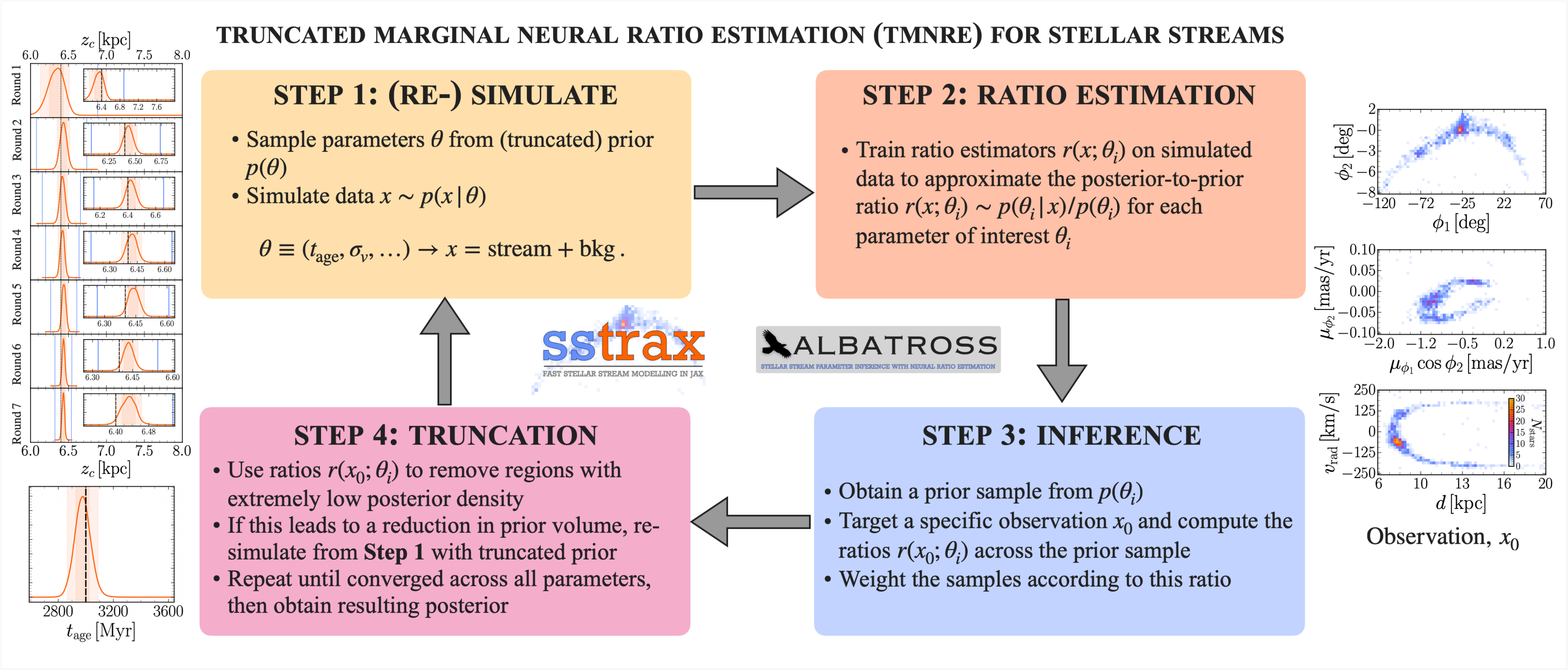}
    \caption{A schematic illustration of the data analysis pipeline developed in this work. We use the TMNRE algorithm (see Sec.~\ref{sec:sbi}) to carry out parameter inference on Milky Way stellar streams (see Sec.~\ref{sec:results}), using our new modelling code \texttt{sstrax} (see Sec.~\ref{sec:stellarstreams}). We also publicly release the \texttt{albatross} analysis code.}\vspace{4pt}
    \label{fig:summary}
\end{figure*}

\noindent \emph{Motivation.} Stellar streams are very old, dynamical objects consisting of a collection of stars that originate from tidal disruptions of a dwarf galaxy (\emph{e.g.} the Sagittarius stream~\cite{Gibbons:2014ewa,Belokurov:2006ms}) or globular cluster (\emph{e.g.} the GD1 stream~\cite{Grillmair:2006bd,Eyre:2009hg,Carlberg:2013gxa,Price-Whelan:2018aaa,Bonaca:2020psc}). In a galaxy such as the Milky Way, these systems have the potential to be an extremely sensitive probe of dark matter substructure~\cite{Erkal:2015kqa,Banik:2018pjp,Banik:2018pal,Banik:2019smi,Banik:2019cza,Bechtol:2019acd,Hermans:2020skz,Malhan:2021aaa,Pavanel:2021wyc}, baryonic physics and Milky Way properties~\cite{Koposov:2009hn,Bonaca:2014qia,Sanderson:2014apa,Amorisco:2016evb,Bovy:2016chl,Bovy:2016irg,Erkal:2019aaa,Helmi:2020otr,Koposov:2023aaa}, as well as the evolution history of the stream~\cite{Balbinot:2017upo,Malhan:2022nfe,Doke:2022jro,Gialluca:2020tno,Banik:2021ymm}. In principle, this can be achieved by combining high precision observations at facilities such as Gaia~\cite{Gaia:2018aaa,Gaia:2021aaa} or the Vera Rubin observatory~\cite{LSSTScience:2009jmu,Bechtol:2019acd}, and consistent modelling of these stellar orbits as the systems are disrupted over the course of $\mathcal{O}$(billions) of years.

\emph{Observational Status.} In the early 2000s, the first observations of cold (and hot) stellar streams in the Milky Way were obtained by the SDSS survey~\cite{SDSS:2008tqn}, the most well-known of which being the GD1 stream~\cite{Grillmair:2006bd,Eyre:2009hg,Carlberg:2013gxa,Price-Whelan:2018aaa,Bonaca:2020psc}. Since then, many more streams have been discovered in surveys such as SDSS and Gaia~\cite{SDSS:2008tqn,Gaia:2018aaa,Gaia:2021aaa,Malhan:2018aab,Ibata:2021aaa,Martin:2022aaa}, but perhaps more importantly, the resolution of the observations has improved dramatically. Current observations have revealed, for example, interesting substructure and features in cold stellar streams such as GD1~\cite{Ibata:2001iv,Johnston:2002aaa,deBoer:2018aaa,Price-Whelan:2018aaa,Bonaca:2020psc}. This is the context in which we want to consistently analyse both the large scale structure of the streams (such as its location on the sky and track), and the small scale structure that is sensitive to \emph{e.g.} the dynamics and details of the tidal stripping process, or baryonic/dark matter interactions.

\emph{Past Analyses.} There are a number of relevant aspects to analysing stellar streams -- stream modelling, inference, and observations. Since the main focus of this work is the statistical analysis of streams, we will briefly review its current status and the corresponding claims, although we will return to computational models for streams when we describe our dynamics code below. Previous analyses have typically focused on either the global structure of the stream, see \emph{e.g.} Refs.~\cite{Bonaca:2014qia,Koposov:2009hn,S5:2021edi,Bowden:2015jva,Gialluca:2020tno,Dillamore:2022aaa,Sanderson:2014apa,Pavanel:2021wyc,Bovy:2016chl,Gibbons:2014ewa} (where it is on the sky and the fits to the general stream track) or construct some sort of summary statistics to study perturbations in the stellar density along the stream object, see \emph{e.g.}~\cite{Bonaca:2018fek,Bovy:2016irg,Banik:2019cza,Erkal:2019aaa,Hermans:2020skz,Banik:2018pjp,Doke:2022jro,Carlberg:2013gxa,Banik:2019smi,Bonaca:2020psc,Amorisco:2016evb,Erkal:2015kqa,Banik:2018pal}.~\footnote{In this regard, the case of GD1 is interesting since there is some evidence that the observed density variations exhibit periodicity along the stream track consistent with the well-known epicyclic variations. See e.g. Fig.~14 in~\cite{Ibata:2019aaa} which constructs the power spectrum as a function of wavenumber along the stream track and highlights a clear peak at $k_s^{-1} \simeq 2.64 \, \mathrm{kpc}$.} The former of these analysis methodologies is well suited for studying properties and phenomena that are specific to the orbit and evolution of a given stream. For example, one can constrain quantities such as the Milky Way potential~\cite{Bonaca:2014qia,Koposov:2009hn,S5:2021edi,Erkal:2019aaa,Bowden:2015jva,Sanderson:2014apa,Craig:2022tcu,Bovy:2016chl,Gibbons:2014ewa,Nibauer:2022aaa,Nibauer:2023aaa}, the age of the stream~\cite{Bovy:2016irg,Hermans:2020skz}, or possibly even gain information about close encounters with large perturbers which can leave large gaps or features in the stream track~\cite{Bovy:2016irg,Banik:2019cza,Banik:2019smi,Bonaca:2020psc,Amorisco:2016evb,Erkal:2015kqa}. The classic examples that are often quoted in the literature along these latter lines are the so-called ``spur" and ``gaps" in the GD1 stream~\cite{Bonaca:2018fek,Doke:2022jro,Carlberg:2013gxa,Price-Whelan:2018aaa}.  On the other hand, the sub-structure of the stream is better suited to asking questions about \emph{e.g.} the physics of tidal stripping mechanisms in the Milky Way, see \emph{e.g.}~\cite{Baumgardt:1998aaa,Baumgardt:2002ya,Taylor:2000zs,Drakos:2022jhg,Takahashi:1999sq}, the internal dynamics and nature of the progenitor, and population level information about smaller (or more distant) perturbers~\cite{Balbinot:2017upo,Gialluca:2020tno,Dillamore:2022aaa,Doke:2022jro,Amorisco:2016evb}. From the perspective of the dark matter community, both the large and small perturbing objects are of huge significance in the context of the distribution of dark matter subhalos in the Milky Way (and other galaxies). Indeed, one of the key goals of stellar stream analyses is to constrain possible subhalo populations~\cite{Banik:2019cza,Hermans:2020skz,Delos:2021ouc,Banik:2018pjp,Pavanel:2021wyc,Banik:2019smi,Banik:2018pal}, or provide a detection of some larger mass (say, $~10^7\,\mathrm{M}_\odot$) subhalo~\cite{Erkal:2015kqa,Bonaca:2018fek}. The main motivation behind our work is to provide a path towards a robust analysis pipeline to consistently (and simultaneously) analyse all of the above scenarios.

\emph{Statistical Challenge.} Making statistically robust statements about quantities of interest -- the gravitational potential of the host, the disruption history, internal dynamics of the progenitor etc. -- can be extremely challenging~\cite{Hermans:2020skz,Huang:2019aaa,Koposov:2023aaa}. To do so requires us to have good control over the dynamical history and initial conditions of the stream~\cite{Bowden:2015jva,Bovy:2014yba,Penarrubia:2005uq,Kuepper:2009sg,Buist:2015hda,Kuepper:2011rx,Fardal:2015aaa,Qian:2022aaa,Bovy:2015aaa}, its stochastic interactions with dark matter or baryonic substructures~\cite{Delos:2021ouc,Bovy:2016irg,Erkal:2014tda}, as well as a reasonable model for foreground and selection effects in the observations, see \emph{e.g.}~\cite{Huang:2019aaa}. As a result of the large number of free parameters this can introduce, together with relatively costly simulations, classical statistical methods scale quite poorly. Currently, this means that one must instead rely on constructing bespoke summary statistics such as the power spectrum of density perturbations along the stream, significantly reducing the dimensionality of the data via \emph{e.g.} only considering the stream track, or ignoring a subset of effects in the modelling to lower the simulation overhead. This approach has been used to obtain relevant results regarding \emph{e.g.} the properties of the Milky Way potential~\cite{Helmi:2020otr,Erkal:2019aaa,Koposov:2023aaa,Sanderson:2014apa,Bonaca:2014qia,Panithanpaisal:2022svk,Gibbons:2014ewa,Koposov:2009hn}, or the evolution history of progenitors~\cite{Doke:2022jro,Banik:2021ymm,Gialluca:2020tno,Malhan:2022nfe,Balbinot:2017upo}. In this paper, we propose using the modern tools and techniques of simulation based inference~\cite{Cranmer:2019eaq,Brehmer:2020cvb} to analyse stellar streams and overcome some of these challenges.

\emph{Simulation Based Inference.} Given the context described above, we briefly argued that the analysis of stellar streams was a problem that is well-suited for the application of simulation-based inference (SBI)~\cite{Cranmer:2019eaq,Brehmer:2020cvb}. Currently, there are a wide range of available approaches and implementations that have been shown to be successful in a number of settings such as CMB data analysis~\cite{Cole:2021gwr}, point source searches~\cite{AnauMontel:2022ppb}, gravitational wave inference~\cite{Bhardwaj:2023xph}, and others, see \emph{e.g.}~\cite{Hermans:2020skz,Montel:2022fhv,Karchev:2022xyn,Gagnon-Hartman:2023soa,Dax:2021tsq}. In general, the advantages of simulation-based inference techniques fall into three categories: \emph{(i)} a consistent inference methodology for any forward simulator, irrespective of the complexity, stochasticity, or data dimensionality of the model, \emph{(ii)} the possibility of extremely simulation efficient inference compared to traditional methods\footnote{This is not necessarily generic across the various methods, but has been observed empirically in a number of settings~\cite{Cole:2021gwr}.}, and \emph{(iii)} the methods do not require an explicit likelihood to be written down, allowing for arbitrarily detailed physics simulations, and observational/detection models. The last point has interesting outlook for stellar streams as it allows for the possibility to significantly improve the modelling and to investigate the implications of \emph{e.g.} selection effects, observation strategies, and instrument errors. This could have important implications for inference results based on \emph{e.g.} small-scale structure in the observed streams or concrete features such as the GD1 spur and gaps~\cite{Bonaca:2018fek,Price-Whelan:2018aaa,deBoer:2018aaa,Carlberg:2013gxa}.

\emph{Key contributions.} This work contributes in a number of ways to the problems and analysis challenges identified above. Firstly, and most importantly, we develop and test a brand new analysis pipeline that leverages recent advances in simulation-based inference. We argue that the use of simulation-based inference to study stellar streams is motivated for a number of reasons. In particular, it allows one to make use of the highest fidelity modelling and observational models via the fact that it is an implicit-likelihood framework. It has also been shown in numerous settings to be highly simulation-efficient~\cite{Cole:2021gwr} compared to more traditional methods such as Markov Chain Monte Carlo (MCMC)~\cite{Mackay:2003aaa,Foreman-Mackey:2012any}.\footnote{In the current context, the application of MCMC techniques to the analysis of streams was pioneered in~\cite{Varghese:2011aaa}.} This is of high relevance to the analysis of streams, since modelling of the complex and varied physics can be computationally costly, making sampling the posterior for large dimensional models typically infeasible. One way we overcome this in the current work is to use a specific targeted (in the sense of analysing a particular observation) simulation-based inference algorithm known as Truncated Marginal Neural Ratio Estimation (TMNRE)~\cite{Miller:2022shs}, implemented within the framework of \texttt{swyft}~\cite{Miller:2021hys,Miller:2022shs}. Secondly, we also developed and will release a public code called \texttt{sstrax} for the modelling of stellar streams in the Milky Way. The current version of the code is designed to be highly modular and extendable for any aspect of streams modelling (\emph{e.g.} the gravitational dynamics or tidal stripping). It is written in \texttt{python} but is highly accelerated through the use of \texttt{jax}~\cite{jax:2018aaa}, allowing for fast ($\mathcal{O}(1)\,\mathrm{s}$) sampling of realistic forward models. This speed is crucial for doing sampling on large dimensional models. Our implementation of the TMNRE algorithm, coupled to the \texttt{sstrax} modelling code will also be made publicly available in the package \texttt{albatross}.

\emph{Structure of the work.} The rest of this work is structured as follows: In Sec.~\ref{sec:stellarstreams} we describe the physics behind the forward modelling of stellar streams, and highlight our numerical implementation in \texttt{sstrax}. Then, in Sec.~\ref{sec:sbi}, we describe the use of simulation based inference for studying and analysing stellar streams, including a detailed explanation of the TMNRE algorithm. In Sec.~\ref{sec:results}, we demonstrate that our analysis pipeline can reliably perform parameter inference on \emph{all} of the parameters in our forward model and discuss the sort of validation tests we can perform on the resulting posteriors. Finally, in Sec.~\ref{sec:conclusions}, we present the key conclusions to the study as well as some outlook as to the relevant use cases and data analysis challenges. 

\section{Modelling Stellar Streams}\label{sec:stellarstreams}

\noindent Arguably one of the most challenging aspects for analysing stellar streams is balancing the complexity of the modelling with the ability to do full parameter inference without resorting to \emph{e.g.} fixing a number of parameters. One of the key arguments we will make later in this work is that simulation-based inference can be a path towards a highly sample efficient analysis framework~\cite{Cole:2021gwr}. This opens up the possibility for using higher fidelity forward models for the dynamics and observation of stellar streams. It is for this reason that we decided to simultaneously develop and test a new modelling code for stellar streams, \texttt{sstrax}, that is modular and designed to be extendable in all aspects with the aim to move towards highly realistic stream modelling for sampling tasks. For the purposes of this work, we have developed what we believe is a simulator that contains all the key elements for for a robust proof-of-principle inference analysis. It will highlight the fact that the analysis and inference pipeline that we develop in later sections is not reliant on particularly symmetric or statistically simple (\emph{e.g.} at the level of the data likelihood) models. We do note, however, that as far as the analysis methodology is concerned, \emph{any} forward model could be used (introducing its own set of modelling assumptions, of course), including \emph{e.g.} the current state-of-the-art models developed in \texttt{galpy}~\cite{Bovy:2015aaa,Bovy:2014yba} or other works~\cite{Erkal:2014tda,Bowden:2015jva,Delos:2021ouc,Bovy:2016irg,Fardal:2015aaa}. 

In this section, we describe the key components to our modelling code, and discuss in each case some relevant improvements that could be made. The generation of a single stream is split broadly into five steps:
\begin{enumerate}[leftmargin=*]
    \item \emph{Cluster Trajectory.} Given some current position $\mathbf{x}_c$ and velocity $\mathbf{v}_c$ for the disrupted cluster, we trace the trajectory back for some time $t_\mathrm{age}$ in the relevant gravitational potential to find the initial conditions.
    \item \emph{Cluster Mass Loss.} We then solve an equation for the evolution of the mass of the cluster $M_c(t)$, due to \emph{e.g.} tidal disruption events, given its trajectory from Step 1, the gravitational potential, and choices for the parameters in the mass loss model.
    \item \emph{Star Stripping Times.} Given this mass loss history, we can then generate a set of stripping times $\{t_i\}_{i = 1..N_\mathrm{stars}}$ for stars released from the cluster. These are chosen to be a random sample from a probability distribution that is a normalised version of $\mathrm{d}M_c/\mathrm{d}t$.
    \item \emph{Stream Stars Evolution.} For each stripping time $t_i$, we generate initial conditions for a star released from the cluster and evolve the star forward in the gravitational potential for a time $(t_\mathrm{age} - t_i)$ before noting its final position and velocity.
    \item \emph{Observation.} Given the full set of stream stars, we construct an observation by projecting to a co-ordinate frame relevant for the stream and accounting for errors in the measurements of \emph{e.g.} the positions and proper motions of the stream stars. We also account for possible background contamination and misidentification that may occur when applying selection cuts.
\end{enumerate}
We will discuss each step in detail below. A concrete example of each step of the analysis process is shown in Fig.~\ref{fig:observation} along with the mock observation used later in the case study.

\subsection{Cluster Trajectory}\label{stellarstreams_trajectory}

\noindent The first step in the modelling is to take the cluster position\footnote{All of our dynamical modelling is performed in a Cartesian co-ordinate frame $(x, y, z)$ with its origin at the galactic centre.} $\mathbf{x}_c = (x_c, y_c, z_c)$ and velocity $\mathbf{v}_c = (v_{x,c}, v_{y,c}, v_{z,c})$ at time $t = 0$ (today) and construct the trajectory for all times $t \in [-t_\mathrm{age}, 0]$. In other words, we project the current position and velocity backwards to find the initial conditions of the cluster a time $t_\mathrm{age}$ ago. To do so, we need to know the gravitational potential $\Phi(\mathbf{x}, t)$ and solve the equation $\ddot{\mathbf{x}}_c(t) = -\nabla \Phi(\mathbf{x}_c, t)$. In terms of implementation, we use the publicly available \texttt{diffrax} differential equation solver library~\cite{Kidger:2021on}, written in \texttt{jax}~\cite{jax:2018aaa}.

In principle, the gravitational potential $\Phi(\mathbf{x}, t)$ can include all contributions from, \emph{e.g.}, the Milky Way dark matter halo, baryonic structures, dark matter subhalos, dynamical clusters or dwarf galaxies etc. In this work, we restrict our attention to a fixed, time-independent Milky Way potential $\Phi = \Phi_\mathrm{MW}(\mathbf{x})$ which consists of a dark matter halo, and baryonic disc and bulge components. Specifically, we choose the \texttt{MWPotential2014} implementation in \texttt{galpy}~\cite{Bovy:2015aaa}, whose parameters are given in Tab.~1 of Ref.~\cite{Bovy:2015aaa}. We note, however, that it is trivial to include arbitrarily complex potentials in our modelling framework. One should also check the level of impact mild to strong mismodelling has in this regards if e.g. the true potential is not exactly the one with which the simulations are generated. One reason for this is that we do not need to analytically construct action-angle co-ordinates (although in principle, this could be possible numerically, see e.g.~\cite{Ibata:2021bbb}). Instead, we take advantage of \texttt{jax}-accelerated differential equation solvers to efficiently evolve the cluster and stars. With this choice, it is also simple to include time-dependent potentials that would arise from either the evolution of the Milky Way itself~\cite{Penarrubia:2005uq,Buist:2015hda,Hammer:2022egx}, or through interactions with dynamical objects such as dark matter subhalos or dwarf galaxies~\cite{Bonaca:2018fek,Bonaca:2020psc,Doke:2022jro,Erkal:2015kqa,Carlberg:2013gxa}. These can be modelled without any additional approximations, and are represented simply by an additional term in the gravitational force. It would also be straight-forward to let the parameters in the Milky Way potential vary and constrain them at the same time as the other model parameters.

\subsection{Cluster Mass Loss}\label{stellarstreams_massloss}

\noindent Once we have the trajectory of the cluster $\mathbf{x}_c(t)$, we want to solve for the evolution of its mass as a function of time $M_c(t)$. This mass loss typically occurs for a number of reasons, due to, \emph{e.g.}, disruption as a result of tidal forces, stellar evolution, or dissolution, see \emph{e.g.}~\cite{Baumgardt:1998aaa,Baumgardt:2002ya,Taylor:2000zs,Drakos:2022jhg,Takahashi:1999sq}. Nonetheless, the vast majority of semi-analytic mass loss models take the form~\cite{Delos:2019lik,vandenBosch:2017ynq,Drakos:2020ksc},
\begin{equation}
    \frac{\mathrm{d}M_c}{\mathrm{d}t} = -\frac{f(M_c, r_t)}{\tau_\mathrm{orb}},
\end{equation}
where $r_t$ is the instantaneous tidal radius, $f$ is some model-dependent function of the cluster mass and tidal radius, and $\tau_\mathrm{orb}$ is some characteristic timescale. In this initial implementation of \texttt{sstrax}, we choose to work with a semi-analytic model given by~\cite{Baumgardt:1998aaa},
\begin{equation}\label{eq:stellarstreams_massloss}
    \frac{\mathrm{d}M_c}{\mathrm{d}t} = -\left(\frac{\xi_0}{t_\mathrm{rh}}\right) \sqrt{1 + \left(\alpha \frac{r_h}{r_t}\right)^3} M_c,
\end{equation}
where $\xi_0$ and $\alpha$ are dimensionless parameters that in initial works were fitted to N-body simulations, $r_h$ is the half-mass radius of the cluster, and $t_\mathrm{rh}$ is the relaxation time given by~\cite{Baumgardt:1998aaa},
\begin{equation}
    t_\mathrm{rh} = 0.138 \, \frac{\sqrt{M_c} \, r_h^{3/2}}{\bar{m}\sqrt{G} \log(0.4 N)}.
\end{equation}
In this expression, $\bar{m}$ is the average mass of a star in the cluster, and $N(t) = M_c(t) / \bar{m}$ is the total number of stars in the cluster. In addition, $r_t$ is the tidal radius, and is computed using~\cite{Bowden:2015jva},
\begin{equation}
    r_t(\mathbf{x}, t) = \left(\frac{G M_c(t)}{\Omega^2 - \mathrm{d}^2 \Phi_\mathrm{MW}/\mathrm{d}r^2}\right)^{1/3},
\end{equation}
where $\Omega$ is the instantaneous angular frequency of the cluster around the galactic centre, $r = |\mathbf{x}|$, and we compute the second derivative of the potential using the auto-differentiation capabilities of \texttt{jax}.

This model quantitatively reproduces interesting features in the mass loss such as the fact that more stars should be stripped near the pericentre of the orbit, which can introduce density variations that are totally separate from \emph{e.g.} epicycles in the stream evolution~\cite{Kuepper:2009sg,Kuepper:2011rx,Ibata:2019aaa}. Again, as in the case of the gravitational potential, this mass loss model can be improved, either by generalising the form, or through a modern calibration to high-resolution N-body simulations of cluster evolution~\cite{Baumgardt:2002ya,Banik:2021ymm,Loyola:2013jta,Rossi:2016aaa,Madrid:2017aaa,Stucker:2022fbn}. Of course, there is a ``gold standard" approach which would be to perform N-body evolution in every simulation. However, we do not expect simulation efficiencies for this type of computation to drop significantly enough for this to become viable in parameter inference. As such, in any inference analysis, one will almost certainly have to resort to a semi-analytic form.

At the level of implementation, we solve this mass loss differential equation numerically using \texttt{diffrax}~\cite{Kidger:2021on}, taking the (densely interpolated) cluster trajectory solution $\mathbf{x}_c(t)$ and initial cluster mass $M_\mathrm{sat}$ as input. As mentioned above, since we directly forward model the mass loss, the code can be modified to use any form of $\mathrm{d}M_c/\mathrm{d}t$, including \emph{e.g.} contributions due to impacts with subhalos or other transient interactions~\cite{Bonaca:2018fek,Bonaca:2020psc,Doke:2022jro,Erkal:2015kqa,Carlberg:2013gxa}.

\subsection{Stripping Times}\label{stellarstreams_stripping}

\noindent Once we have obtained the cluster mass $M_c(t)$ as a function of time $t$, we want to stochastically generate a set of stripping times $\{t_i\}$. These times define the moment the stars which will ultimately form the final stream are released.
To go from cluster mass to stripping times, we identify $(-\mathrm{d}M_c(t)/\mathrm{d}t)$ as the instantaneous stripping rate. We could model this faithfully as an inhomogeneous Poisson process, however a simple approximate scheme, which we outline below, is sufficient for our purposes.

First, we introduce the average mass of a star $\bar{m}$ as a new parameter, although we note that this is a somewhat toy simulation parameter since real systems are known to not have monochromatic mass functions.
We then compute the total number of stars that should be in the final stream as $N_\mathrm{stars} = \Delta M/\bar{m}$, where $\Delta M = (M_\mathrm{sat} - M_c(t = 0))$ is the total mass loss of the cluster.\footnote{For context, in a stream such as GD-1, there are typically around 1000 stars reported as probable members (with some variation according to the detection method). In the mock observation we present here, there are $N_\mathrm{stars} = 968$ stars, so represents qualitatively the same scale of system.}
Now, each of the $N_\mathrm{stars}$ stripping time can be sampled individually according to the distribution $(-\mathrm{d}M_c(t)/\mathrm{d}t) / \Delta M$. This can be done in a number of ways, but in \texttt{sstrax} we choose to construct the cumulative distribution function and sample uniformly from $U[0, 1]$ before projecting back to the $t$-space.

Note that this scheme does not explicitly use a distribution over star masses.
Nevertheless, if one were to compute the differences of the heuristic cluster mass function between stripping times, one would obtain some mass distribution clustered around $\bar{m}$.
The important point to realise is that we already make an approximation in using a continuous cluster mass $M_c(t)$. If the particular distribution of star masses becomes relevant, both the stripping and the cluster mass modeling would have to be replaced by a more realistic framework.
This could be achieved, for example, by sampling the next star mass $m_s \sim p(m_s)$~\cite{Schulz:2015aaa}, treating the cluster mass loss in Eq.~\eqref{eq:stellarstreams_massloss} as the instantaneous event frequency of an inhomogeneous Poisson process, and only reducing the cluster mass by discrete steps $m_s$ whenever a star is released.
It would be interesting to explore the implications of these two effects (i.e. star mass distributions and modelling the cluster mass via an instantaneous Poisson process instead of a continuous function) in the context of limits derived from density variations along the stream tracks, see \emph{e.g.}~\cite{Banik:2018pjp,Banik:2019cza,Hermans:2020skz}. We have also assumed that the cluster system is collisionless,
whereas in some systems populations of e.g. dark masses (see e.g.~\cite{Vitral:2022apu}) towards their centre or
accretion of halo clusters (see~\cite{Mackey:2019aaa,Malhan:2019aaa}) could break this assumption and should
be modelled properly before targeting real data.

Finally, note that our stripping process is stochastic and can therefore lead to different realisations of the density profile along the stream track if the same stream is generated multiple times. 

\subsection{Stream Star Evolution}\label{stellarstreams_evolution}

\noindent The final dynamical step for generating the stream is quite simple -- we just need to release stars from nearby the cluster at the times $t_i$ and evolve them forward in the same gravitational potential $\Phi(\mathbf{x}, t)$\footnote{In principle, one can also evolve them in the gravitational potential of the cluster as well as the Milky Way, but we found that this was indistinguishable at the level of inference results. It is also well-known that we do not need to include the self-gravity of the stream itself~\cite{Delos:2021ouc}.} as the cluster until today $t = 0$. The only choice left to be made is one regarding the initial conditions for the stars, which we choose in accordance with observations made in N-body simulations of tidally disrupted clusters~\cite{Baumgardt:2002ya,Loyola:2013jta,Madrid:2017aaa,Stucker:2022fbn,Rossi:2016aaa}. It has been shown that the majority of stars escape from near one of the two Lagrange points $\mathbf{x}_{1,2} = (1 \pm (r_t/r))\mathbf{x}_c$ of the cluster~\cite{Varghese:2011aaa,Bowden:2015jva} (one on either side of the radial line joining the galactic centre and the cluster centre), where $r = |\mathbf{x}_c|$.

In the \texttt{sstrax} implementation, we generalise this slightly and introduce three additional parameters: $\lambda_{\mathrm{rel}}$, $\lambda_{\mathrm{match}}$, and $p_\mathrm{near}$. Respectively, these describe how far away from the cluster the star is released, i.e. $\mathbf{x}_\mathrm{rel} = (1 \pm \lambda_{\mathrm{rel}}(r_t/r))\mathbf{x}_c(t_i)$, at what distance the velocity matching is done (specifically, the velocity is matched so that the angular velocity of the star and the cluster agree at a distance $\mathbf{x}_\mathrm{match} = (1 \pm \lambda_{\mathrm{match}}(r_t/r))\mathbf{x}_c(t_i)$), and finally the probability $p_\mathrm{near}$ of being released from the closer Lagrange point. Finally, to model the velocity dispersion of the cluster itself, we choose the initial velocity of the star to be this matching velocity plus an additional random vector $\Delta \mathbf{v}$ sampled on the unit sphere and rescaled by a factor $\sqrt{3} \sigma_v$, where $\sigma_v$ is the velocity dispersion. 

In much the same way as the mass loss model, the most realistic way to actually model this process would be to account for the full dynamics inside the cluster via some N-body approach. For the same reason, this is still too costly for parameter inference tasks, so a semi-analytic approach like the one above needs to be used. Again, and in line with the prescription we chose for the mass loss, since we directly forward model the evolution of the stars, the generation of these initial conditions can be tuned arbitrarily to either analytic expectations, or some new high-resolution simulations. In any case, the analysis pipeline will remain the same.

\subsection{Observational Model}\label{stellarstreams_observation}

\noindent An important aspect of simulation-based inference approaches is that the forward model \emph{must} also include the detector response, observational model, or noise generation. This is in contrast perhaps to traditional approaches where typically some clean signal output of the forward model is input into an explicit data likelihood. In practice, the statistics results should be identical in either formulation. In the context of stellar streams, given some final stream configuration $\{\mathbf{x}_\star^i, \mathbf{v}_\star^i\}_{i = 1...N_\mathrm{stars}}$, we need to model, \emph{(a)} the observational measurement errors, \emph{(b)} the detection of the stream in the sky, and \emph{(c)} the contaminating background of other stars.

In this work, we develop a simple initial observational model, meant mostly as a proof of principle. Specifically, we assume that the stream has been ``detected" through some form of selection cuts and vetoes in survey data~\cite{Huang:2019aaa,Shih:2021kbt,Malhan:2018aaa,Borsato:2020aaa,Shih:2023jfv}. We use this to define an observational window which we choose to focus on (i.e. we do not model the full sky). In the rest of the analysis, we will be focusing on a mock stream that is supposed to resemble the GD1 stream~\cite{Price-Whelan:2018aaa,Grillmair:2006bd,Eyre:2009hg}. There are a standard set of co-ordinates used in the literature~\cite{Koposov:2009hn} to describe the phase-space structure of this stream. Specifically, there are two angle co-ordinates $(\phi_1, \phi_2)$ which are approximately aligned with the stream track at $\phi_2 \simeq 0\,\mathrm{deg}$, the corresponding proper motions $(\mu_{\phi_1}, \mu_{\phi_2})$, and radial distances and velocities $(d, v_\mathrm{rad})$. The definitions for these can be found in Appendix~\ref{app:sstrax}.

Given these definitions, to construct the observation from the list of stream stars, we first define the region of interest in the sky/velocity phase space, i.e. we ignore all stars with $(\phi_1, \phi_2, ..., v_\mathrm{rad}) \notin [\phi_1^\mathrm{min}, \phi_1^\mathrm{max}] \times \cdots \times [v_\mathrm{rad}^\mathrm{min}, v_\mathrm{rad}^\mathrm{max}]$. Then, we add random observational errors to the values generated by \texttt{sstrax} via sampling \emph{e.g.} $\phi_1^\mathrm{obs} \sim \mathcal{N}(\phi_1, \delta\phi_1)$. Finally, we model two aspects of stream detection and selection effects. In particular, we assume that we have some selection efficiency $\epsilon_\mathrm{sel}$ that measures how often we accidentally miss a star in a given detection algorithm that should have been correctly classified as part of the stream~\cite{Huang:2019aaa,Shih:2021kbt,Malhan:2018aaa,Borsato:2020aaa,Shih:2023jfv}. We also model the fact that there can be contamination from the background stars that are not part of the stream, but are nonetheless not removed by the detection algorithm and are in the observing window~\cite{Huang:2019aaa}. This is quantified by assuming there is some number $N_\mathrm{background}$ stars, of which we are able to successfully remove $(1 - \epsilon_\mathrm{background})$\% via the selection process. We then distribute $N_\mathrm{background} \epsilon_\mathrm{background}$ stars uniformly across the observational windows to model the background contamination. Finally, we bin the remaining data into three channels of size $(N_\mathrm{bins}^x, N_\mathrm{bins}^y)$ each: $(\phi_1, \phi_2)$, $(\mu_{\phi_1}, \mu_{\phi_2})$, and $(d, v_\mathrm{rad})$.\footnote{It is worth noting that for streams with relatively low stellar counts, binning the data may not be the most appropriate data representation. Arguably one of the key benefits of the simulation-based inference paradigm, however, is that if a more relevant data choice can be made/simulated, then the statistical implications will be automatically taken into account. This final point also holds if \textit{e.g.} some aspects of the data are unavailable for some reason (such as the radial positions and velocities in the current context).} All the choices for the particular values of the observational model described here are given in Tab.~\ref{tab:obs_parameters}.

As in the other components, there is significant room for more detailed modelling. For example, we know just from looking at \emph{Gaia} data that the background stars will not be uniformly distributed across the sky~\cite{Gaia:2018aaa,Gaia:2021aaa,Boubert:2020aaa}, with higher concentrations near the galactic centre. Similarly, the efficacy of the sort of selection criteria or cuts that are applied based on \emph{e.g.} metallicity or proper motions are likely at least stream- and sky location-dependent. The extent to which this impacts the inference is a different question, and something that we can actually test in our framework by modifying the observational model.

\subsection{Acceleration with \texttt{jax}}\label{stellarstreams_jax}

\noindent Making the decision to directly forward model the evolution of the stream, rather than construct either some effective description~\cite{Delos:2021ouc}, or accelerate the dynamical solutions through action-angle co-ordinate constructions opens up the possibility for far more general simulation frameworks. On the other hand, it is also potentially much more computationally intensive, \emph{e.g.} if we include the effect of a large population of subhalos in the future. This is compounded by the additional simulation budget that is potentially required to perform inference on the large number of parameters any augmentation of the model can introduce. 

As such, an important component of our implementation is its computational efficiency. We have achieved this by using the \texttt{jax} framework~\cite{jax:2018aaa}, which allows for just-in-time compilation caching and highly optimised custom vectorisation.

\begin{figure*}[t]
    \centering
    \includegraphics[width=\linewidth,trim={0.1cm 0.1cm 0.1cm 0.1cm},clip]{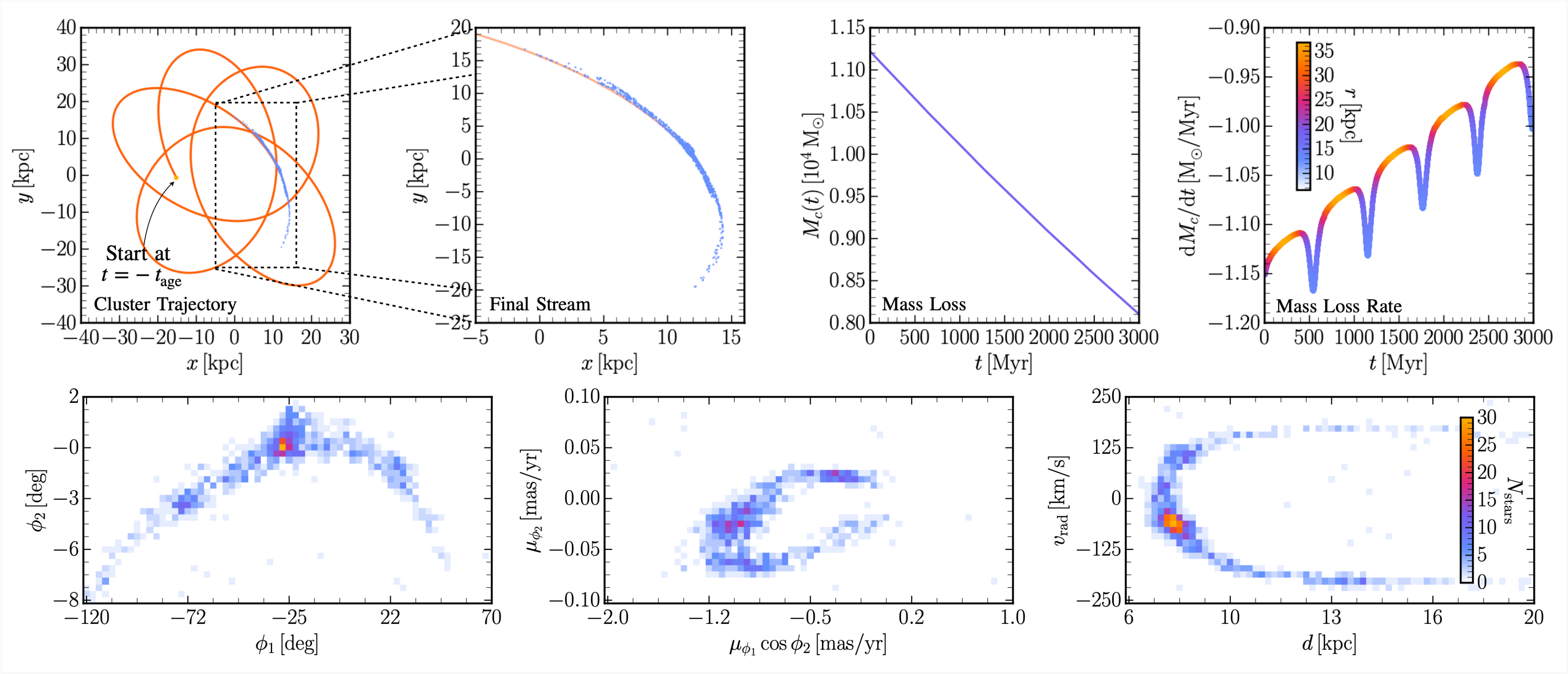}
    \caption{\emph{Upper panels:} Illustration of the various modelling steps (finding the cluster trajectory, mass loss history, final stream formation etc.) described in Sec.~\ref{sec:stellarstreams}. \emph{Lower panels:} Example mock observation generated using the \texttt{sstrax} modelling code with the parameters in Tab.~\ref{tab:parameters}. Analysed as a case study in Sec.~\ref{sec:results}.}
    \label{fig:observation}
\end{figure*}

\section{Simulation Based Inference for Stellar Streams}\label{sec:sbi}

\noindent In this section, we will give a brief review of general simulation-based inference (SBI) methods before describing the specific implementation we will use in this work. We will end the section by presenting some of the algorithm design choices that are relevant to stellar stream analysis.

\subsection{Overview of Simulation-based Inference}\label{sec:sbi_overview}

\noindent Recently, there have been significant advances in high fidelity physics simulations, and machine learning techniques for processing complex data structures, alongside the emergence of increasingly challenging data analysis problems. This has led to the rapid development of ``simulation-based inference" as a competitive alternative to traditional techniques as far as scalability, model realism, and unbiased analysis pipelines are concerned~\cite{Cranmer:2019eaq,Brehmer:2020cvb}. At its heart, the field of simulation-based inference asks: \emph{given some forward model or simulator, can we perform efficient and correct Bayesian inference?} Ultimately, the goal of simulation-based inference is to develop a robust statistical pipeline that can make use of the most realistic and state-of-the-art modelling tools.

To be more concrete, suppose we have some forward model $p(x, \theta)$ that takes the model parameters $\theta$ -- which could be a range of physical parameters, effective model components, nuisance parameters etc. -- to some data $x$ that resembles the real observed data $x_0$. In a Bayesian context, we sample $\theta$ from some chosen\footnote{Ideally with some physical motivation for the ranges chosen, or some maximally uninformative choice otherwise.} prior distribution $p(\theta)$ so that the forward model takes the form $p(x, \theta) = p(x | \theta) p(\theta)$. This expression is at the heart of simulation-based methods, since it formally represents the notion that ``running your simulator" is the same as sampling from the (simulated-)data likelihood $p(x | \theta)$. Indeed, this is the origin of the terms ``likelihood-free" or ``implicit likelihood" inference to describe SBI~\cite{Cranmer:2019eaq,Brehmer:2020cvb}. These descriptions are supposed to convey the distinction between analytically evaluating some expression to compute the likelihood $p(x | \theta)$ and sampling from it.

To understand the different ways in which SBI methods approach the Bayesian inference problem, it is useful to briefly review how the forward model fits into Bayes' theorem. As far as scientific conclusions are concerned, we are typically\footnote{Of course, there are use cases \emph{e.g.} in model comparison or goodness-of-fit tests~\cite{SpurioMancini:2022vcy}, where computing other quantities such as the data evidence or maximum likelihood is more relevant.} interested in computing the posterior $p(\theta | x)$ of the parameters given some data $x$,
\begin{equation}\label{eq:sbi_bayes}
p(\theta | x) = \frac{p(x | \theta) p(\theta)}{p(x)}.
\end{equation}
Here, as above, $p(x | \theta)$ is the data likelihood, $p(\theta)$ is the prior over our parameters $\theta = (\theta_1, \cdots)$, and $p(x)$ is the evidence. Given this setup, there are various ways that SBI algorithms tackle posterior estimation given the ability to sample from the forward model $(x, \theta) \sim p(x, \theta) = p(x | \theta) p(\theta)$. Specifically, these can be categorised as follows (with the method we use highlighted in bold):
\begin{itemize}[leftmargin=*]
    \item \emph{Neural Posterior Estimation (NPE).} In neural posterior estimation~\cite{Papamakarios:2016aaa,Zeghal:2022aaa}, the goal is to directly estimate the posterior distribution $p(\theta | x)$ by representing it as some flexible parameterised probability density. This has been applied successfully in a number of contexts, \emph{e.g.} gravitational wave analysis~\cite{Dax:2021tsq} and open source implementations are available~\cite{Tejero-cantero:2020aaa}.
    \item \emph{Neural Likelihood Estimation (NLE).} In contrast, neural likelihood estimation~\cite{Alsing:2019xrx,Papamakarios:2019aaa} attempts to construct an estimator for the (simulated-)likelihood function itself $p(x | \theta)$. This can then be used to carry out standard inference techniques such as MCMC~\cite{Mackay:2003aaa,Foreman-Mackey:2012any} or nested sampling~\cite{Ashton:2022grj,Skilling:2006gxv,Handley:2015aaa} and generate samples from the posterior.
    \item \emph{\textbf{Neural Ratio Estimation (NRE).}} Finally, neural ratio estimation~\cite{Hermans:2019aaa,Durkan:2020aaa,Miller:2022shs,Delaunoy:2022aaa,Rozet:2022aaa} considers the ratio $p(x | \theta) / p(x)$ appearing on the right hand side of Eq.~\eqref{eq:sbi_bayes}. This particular approach will be the focus of this work, in the form of an algorithm known as Truncated Marginal Neural Ratio Estimation (TMNRE)~\cite{Miller:2022shs}, implemented within the framework of \texttt{swyft}~\cite{Miller:2021hys}.
\end{itemize}

\subsection{The TMNRE Algorithm}\label{sec:sbi_tmnre}

\noindent We will now focus on the specific implementation of simulation-based inference used in this work. This is known as Truncated Marginal Neural Ratio Estimation (TMNRE)~\cite{Miller:2022shs}, and is implemented in the \texttt{swyft} software~\cite{Miller:2021hys}. We have summarised the method in Fig.~\ref{fig:summary} for reference, however, there are a number of features we wish to emphasise in terms of its applicability to stellar streams.
\begin{itemize}[leftmargin=*]
    \item \emph{Targeted Inference.} TMNRE is both a ``targeted" and ``sequential" algorithm in the sense that it performs inference on a specific target observation $x_0$ (as opposed to amortising over all possible model outputs) over a number of discrete rounds. In each round, the prior is truncated based on inference in the current round (see description below) to avoid simulating in parameter regions which do not contribute significantly to the likelihood ratio, given the fixed target observation.
    \item \emph{Marginal Posteriors.} There are a number of quantities of interest in Bayesian inference, including the full joint posterior $p(\theta | x_0)$ given some observation $x_0$, the evidence of a particular observation $p(x_0)$, or marginalised posteriors\footnote{In the strict technical sense that $p(\theta_i^\star | x) = \int{\mathrm{d}^n \theta \, p(\theta | x) \delta(\theta_i - \theta_i^\star)}$} $p(\theta_i | x_0)$ for some individual parameter $\theta_i$ or subset of parameters in $\theta$. In TMNRE, a significant portion of the achieved simulation efficiency arises due to the fact that we directly estimate the marginal posterior, rather than marginalising over samples from the full joint distribution.
\end{itemize}
In combination, these two properties are the key to achieving a highly simulation efficient inference strategy. For more discussions along these lines, see \emph{e.g.} works discussing the application of TMNRE to CMB~\cite{Cole:2021gwr} and gravitational wave analyses~\cite{Bhardwaj:2023xph}.

Although the details of the method can be found in the original literature~\cite{Miller:2022shs}, it is useful to give a brief overview of the setup of the ratio estimation problem. This will highlight the features described above and how they will be beneficial for the analysis of stellar streams. The goal of TMNRE is to estimate the following ratio,
\begin{equation}\label{eq:sbi_ratio}
    r(x; \theta) = \frac{p(x | \theta)}{p(x)} = \frac{p(\theta | x)}{p(\theta)} = \frac{p(x, \theta)}{p(x)p(\theta)},
\end{equation}
where the last two equalities follow from an application of Bayes' theorem in Eq.~\eqref{eq:sbi_bayes} and the definition of the joint distribution $p(x, \theta) \equiv p(x | \theta)p(\theta)$. In other words, access to the ratio $r(x; \theta)$ is equivalent to estimating \emph{(a)} the likelihood-to-evidence ratio $p(x | \theta)/p(x)$, \emph{(b)} the posterior-to-prior ratio $p(\theta | x)/p(\theta)$ which will be used for parameter estimation, and \emph{(c)} the joint-to-marginal ratio $p(x, \theta)/p(x)p(\theta)$ which will be the technically important form to perform ratio estimation in practice.

If we focus on the last form, $r(x; \theta) = p(x,\theta)/p(x)p(\theta)$, we can make the observation that given a set of simulations $\{(x, \theta)\}$ from our forward model $p(x, \theta) = p(x | \theta)p(\theta)$, we can construct two distinct classes of sample. The first is simply a sample from the full joint distribution $p(x, \theta)$, which amounts to picking an individual simulation pair $(x, \theta)$. The second is a sample from the combined marginal distribution $p(x)p(\theta)$ which can be obtained by picking two random samples, then taking $x$ from one, and $\theta$ from the other. Having these two distinct distributions is the origin of ratio estimation as a binary classification task\footnote{This can actually be generalised in interesting ways to form multi-class classification problems that are applicable to \emph{e.g.} hierarchical models~\cite{Miller:2022haf}.} -- it asks the question \emph{given a pair $(x, \theta)$, did $\theta$ generate $x$?} Intuitively, the relative precision of the posterior distribution in this case reflects how difficult it is to discriminate between joint and marginal samples. For instance, the larger the observational error is, the more overlap there will be between these two classes and the posterior will be wider.

More formally, we can frame this binary classification task and ratio estimation as an attempt to optimise (specifically minimise) the following loss function\footnote{This is nothing other than the binary cross-entropy for a classifier that tries to discriminate between joint $(x, \theta) \sim p(x, \theta)$ and marginal $(x, \theta) \sim p(x)p(\theta)$ samples.}~\cite{Miller:2021hys},
\begin{multline}\label{eq:sbi_loss}
    \mathcal{L}[f_\phi] = - \int{\mathrm{d}x\mathrm{d}\theta} \, p(x, \theta) \ln \left(\sigma(f_\phi(x, \theta)) \right) \\ + p(x)p(\theta)\ln \left(1 - \sigma(f_\phi(x, \theta)) \right).
\end{multline}
Here $\sigma(x) = [1 + \exp(-x)]^{-1}$ is the sigmoid function, and $f_\phi(x, \theta)$ is the classifier with some set of free parameters $\phi$ that should be optimised. One of the key justifications for the correctness of TMNRE as an inference algorithm is to realise that this loss can actually be minimised analytically. In particular, one can show that the optimal classifier is given by $f_\phi^\star(x, \theta) = \ln r(x; \theta)$~\cite{Miller:2021hys}. In other words, if one can successfully minimise the loss in Eq.~\eqref{eq:sbi_loss}, then \emph{one directly obtains the posterior-to-prior ratio $r(x; \theta)$}.

Practically, this is where the ``N(eural)'' part of TMNRE is relevant, especially for very high dimensional data/parameter spaces. Modern machine learning methods, architectures, and hardware allow for very flexible parameterisations of the classifier $f_\phi$, and there is a well established methodology to optimise their parameters $\phi$ (also more commonly called their weights). As far as the analysis of stellar streams is concerned, this gives us access to data representations that are as close as possible to real data from \emph{e.g.} Gaia, without any need for compression into hand-crafted summary statistics, spline fits, or similar data reductions. In this work, the task of optimising is achieved through the use of the software \texttt{swyft}, which is built on top of \texttt{pytorch}.

With (neural) ratio estimation set up this way, we can now see how to directly estimate marginal posteriors in this framework. Suppose we wish to estimate the fully marginalised posterior for a single parameter\footnote{Or any subset of parameters $(\theta_{i_1}, \theta_{i_2}, \ldots, \theta_{i_k})$ to generate the $k$-dimensional marginal posterior $p(\theta_{i_1},\ldots,\theta_{i_k} | x)$. We will give explicit examples for $k=2,3$ in Sec.~\ref{sec:results}.} $\theta_i$ in $\theta$, then we can start by taking our full suite of simulations $(x, \theta) \sim p(x, \theta)$ which (crucially) vary \emph{all} parameters $\theta$. Then, however, instead of constructing the loss based on all parameters, we can only ``show" the single parameter $\theta_i$. This is equivalent to replacing $\theta \rightarrow \theta_i$ in Eq.~\eqref{eq:sbi_loss} above. Importantly, however, the analytic arguments will still hold and allow us to obtain directly the marginal posterior-to-prior ratio $p(\theta_i | x)/p(\theta_i)$. In contrast to \emph{e.g.} MCMC where this marginalisation is performed \emph{after} obtaining samples from the posterior, we implicitly marginalise by varying all parameters in the simulations simultaneously, but constructing the marginal posterior directly rather than via the joint. As far as analysing stellar streams is concerned, this is not just a useful trick for quickly obtaining the marginal posteriors, but is crucial in making the algorithm simulation efficient. Looking forwards, if the goal is to perform inference with extremely high fidelity stream simulations in order to extract the maximum possible information from the data, analysis methods that break the traditional scaling of sampling algorithms such as MCMC or nested sampling will be vital.

The final aspect to discuss before we summarise the algorithm and the design choices relevant to stellar streams is the truncation process that allows us to target a particular observation $x_0$. As far as simulation efficiency is concerned, the idea behind truncation is to minimise the number of simulations performed in regions where there is extremely low posterior density, since, by definition, they provide almost no information about the parameter estimation problem. Formally we achieve this by performing the inference sequentially in several rounds. In each round, we generate a set of simulations $(x, \theta) \sim p(x, \theta)$ from the full model. Then we train and optimise our classifiers $f^i_\phi(x, \theta_i)$ for the parameters of interest from which we can obtain marginal posteriors on each parameter $p(\theta_i | x_0)$ for some specific target observation $x_0$. This will highlight regions of parameter space where the posterior density for $\theta_i$ is both very high, and of course other regions where the density is low, indicating that given the observation $x_0$, this particular set of parameters is unlikely. We use these latter regions to truncate our prior region by imposing the condition $r_i(x_0, \theta_i) < \epsilon$ on the estimated ratios\footnote{Of course, this will introduce a slight error in the estimate of the marginal posterior proportional to $\Delta p(\theta^\star_i | x_0) \sim \int_{\Gamma(\epsilon)}{\mathrm{d}^n \theta \, p(\theta | x_0) \delta(\theta_i - \theta_i^\star)}$, where $\Gamma(\epsilon)$ is the region excluded by the truncation procedure. However, it is exactly in this region where the joint posterior density is (necessarily) low, and as such, the error induced is small and strictly controlled by $\epsilon$. To be conservative, we typically choose $\epsilon \sim 10^{-5}$, which corresponds to exclusion at around the $4.5\sigma$ level for a Gaussian distribution~\cite{Miller:2022shs}. Provided $\epsilon$ is not too large, any other choice should not change our results at all, only affecting the time that the algorithm takes to converge.}. Then, we re-simulate by sampling from this truncated prior, repeat the inference and then truncate again. Eventually, once the posteriors converge to the level of statistical uncertainty, the truncation will just return the restricted prior and the algorithm will terminate. This truncation process is highlighted below in Fig.~\ref{fig:truncation}.

In summary, the TMNRE algorithm splits into four steps that are highlighted in the schematic shown in Fig.~\ref{fig:summary}:

\begin{enumerate}
    \item \textbf{Step 1:} Sample a set of simulations\footnote{Note that this step can be fully parallelised, something that is implemented directly in \texttt{albatross}.} from the full forward model $(x, \theta) \sim p(x, \theta) = p(x | \theta) p(\theta)$.
    \item \textbf{Step 2:} Train a set of classifiers $f^i_\phi(x, \theta_i)$ to obtain an estimate of the ratio $r_i(x; \theta_i) = p(\theta_i | x)/p(\theta_i)$.
    \item \textbf{Step 3:} Use this trained ratio to obtain estimates of the marginal posteriors $p(\theta_i | x_0)$ for a specific target observation $x_0$.
    \item \textbf{Step 4:} Take these marginal posterior distributions and derive bounds on the prior region to truncate for the next round of inference by imposing the condition $p_i(\theta_i | x_0)/ \mathrm{max}_{\theta_i}p_i(\theta_i | x_0) < \epsilon$.
    \item Repeat from \textbf{Step 1} until the truncation procedure stabilises, then take the final round of inference as the set of posteriors $p(\theta_i | x_0)$ and terminate the algorithm.
\end{enumerate}

\begin{table}[t]
\centering
\resizebox{\linewidth}{!}{
\begin{tabular}{@{}lll@{}}
\hline
\textbf{Parameter}                                                         & \textbf{Prior Range}                            & \textbf{True Value}\\ \hline
(Log of) Initial Cluster Mass $\log_{10}(M_\mathrm{sat}/\mathrm{M}_\odot)$ & $[3.0, 4.5]$                                    & $4.05$\\
Cluster Velocity Dispersion $\sigma_v$                                     & $[0.1, 5.0] \, \mathrm{km} \, \mathrm{s}^{-1}$  & $1.1$ \\
Cluster Final Pos. $\mathbf{x}_\mathrm{sat} = (x_c, y_c, z_c)$             & Stream Dependent${}^\star$                      & $(11.8, 0.79, 6.4)$ \\
Cluster Final Vel. $\mathbf{v}_\mathrm{sat} = (v_{x,c}, v_{y,c}, v_{z,c})$ & Stream Dependent${}^\star$                      & $(109.5, -254.5, -90.3)$ \\
Stream Age $t_\mathrm{age}$                                                & $[500, 5000] \, \mathrm{Myr}$                   & $3000$ \\
Release Distance Parameter $\lambda_\mathrm{rel}$                          & $[0.1, 2.0]$                                    & $1.405$ \\
Release Velocity Parameter $\lambda_\mathrm{match}$                        & $[0.1, 2.0]$                                    & $1.846$ \\
Stripping asymmetry $p_\mathrm{near}$                                      & $[0, 1]$                                        & $0.5$ \\
Mass loss prefactor $\xi_0$                                                & $[10^{-4}, 10^{-2}]$                            & $0.001$ \\
Mass loss parameter $\alpha$                                             & $[10, 30]$                                      & $20.9$ \\
Half-mass Radius $r_h$                                                     & $[10^{-4}, 10^{-2}] \, \mathrm{pc}$             & $0.001$ \\
Average Stellar Mass $\bar{m}$                                             & $[1.0, 20] \, \mathrm{M}_\odot$                 & $3$ \\ \hline
\end{tabular}}
\caption{Parameters, prior range choices and injection values for the stream model parameters described in Sec.~\ref{sec:stellarstreams}. \\
{\footnotesize ${}^\star$In particular, we choose these parameters to span the observational window of interest for an individual observation. In the analysis presented in Sec.~\ref{sec:results}, we choose the priors $([10, 14], [0.1, 2.5], [6, 8])$ and $([90, 115], [-280, -230], [-120, -80])$ on the cluster position and velocity respectively.}}
\label{tab:parameters}
\end{table}

\subsection{Design Choices for Stellar Streams}\label{sec:sbi_choices}

\noindent In order to use the TMNRE algorithm in practice, we must make a number of design choices. These include \emph{(a)} building or using a pre-implemented forward model that generates the data $x$ (here a representation of the stellar stream) given parameters $\theta$, \emph{(b)} designing a neural network architecture that is able to efficiently process the data format of $x$ and $\theta$, \emph{(c)} making choices for the prior distributions over the parameters $\theta$, and \emph{(d)} choosing the hyperparameters relevant to the TMNRE algorithm.

\vspace*{8pt}
\noindent \emph{Forward Simulator.} To generate stellar stream simulations, we use the implementation of our modelling approach described in detail above (see Sec.~\ref{sec:stellarstreams}). To very briefly recap, we solve for the full evolution history of the stream including \emph{e.g.} the orbit-dependent tidal stripping in a framework that can accommodate for any time-dependent or time-independent gravitational potential. In addition, we develop a simple observational model that is supposed to represent experimental and statistical uncertainties at the level of a current survey. This is implemented in the \texttt{jax}-accelerated modelling code \texttt{sstrax}, which we couple directly to the \texttt{swyft} software~\cite{Miller:2021hys} in our analysis code \texttt{albatross}. The parameters that we vary in this analysis $\theta = (t_\mathrm{age}, M_\mathrm{sat}, \ldots)$ are described in Tabs.~\ref{tab:parameters} (stream modelling) and~\ref{tab:obs_parameters} (observation model). An example output of our simulator (and the case study that we investigate below) is shown in Fig.~\ref{fig:observation}.

\begin{table}[t]
\centering
\resizebox{\linewidth}{!}{
\begin{tabular}{@{}ll@{}}
\hline
\textbf{Observation Model Parameter} & \textbf{Value} \\ \hline
Observing window $\phi_1$ & $[-120, 70]\,\mathrm{deg}$ \\
Observing window $\phi_2$ & $[-8, 2]\,\mathrm{deg}$ \\
Observing window $\mu_{\phi_1}\cos\phi_2$ & $[-2, 1]\,\mathrm{mas/yr}$ \\
Observing window $\mu_{\phi_2}$ & $[-0.1, 0.1]\,\mathrm{mas/yr}$ \\
Observing window $d$ & $[6, 20]\,\mathrm{kpc}$ \\
Observing window $v_\mathrm{rad}$ & $[-250, 250]\,\mathrm{km/s}$ \\
Number of bins & $[64, 32]$ \\
Observational error $\delta \phi_1$ & $0.001\,\mathrm{deg}$ \\
Observational error $\delta \phi_2$ & $0.15\,\mathrm{deg}$ \\
Observational error $\delta \mu_{\phi_1}\cos\phi_2$ & $0.1\,\mathrm{mas/yr}$ \\
Observational error $\delta \mu_{\phi_2}$ & $0.0\,\mathrm{mas/yr}$ \\
Observational error $\delta d$ & $0.25\,\mathrm{kpc}$ \\
Observational errors $\delta v_\mathrm{rad}$ & $5\,\mathrm{km/s}$ \\
Stream selection success rate $\epsilon_\mathrm{sel.}$ & $95\%$ \\ 
Background stars in window $N_\mathrm{background}$  & $10^6$ \\
Background contamination rate $\epsilon_\mathrm{background}$ & $10^{-3}\%$ \\ \hline
\end{tabular}
}
\caption{Choices for observational model parameters described in Sec.~\ref{sec:stellarstreams}.}
\label{tab:obs_parameters}
\end{table}

\vspace*{8pt}
\noindent \emph{Inference Network.} As discussed above, the main aim of (neural) ratio estimation is to design a procedure that can reliably train a classifier, or set of classifiers $f^i_\phi(x, \theta_i)$ to distinguish between joint and marginal samples~\cite{Miller:2021hys,Durkan:2020aaa,Rozet:2022aaa,Delaunoy:2022aaa,Hermans:2019aaa}. To do this in practice, we need a flexible way to parameterise $f_\phi$, and although there are arguments from \emph{e.g.} the loss function in Eq.~\eqref{eq:sbi_loss} that any flexible enough parameterisation (i.e. just having enough trainable parameters in $\phi$) will be able to optimise the loss, in reality this is only in some infinite training data limit. As such, we should try to make sensible design choices regarding the network to take full advantage of the known structure and physics associated to our signal and data format. Empirically, making physics-informed choices at this stage leads to huge increases in performance, robustness, simulation efficiency, and the general applicability of the method.

In our case of stellar streams, the signal is a collection of stars and their properties (positions, velocities, perhaps even metallicity etc.). As described in Sec.~\ref{sec:stellarstreams}, we focus on the phase-space information in this work, including \emph{e.g.} the selection procedure in our observational model. As part of our forward model, we chose to bin the data into a 3-channel image to preserve the spatial structure and morphology of the signal. This data format is then well suited for the application of standard image processing network structures.

More precisely, we know that a stream binned at a given resolution can have structure on a range of scales. For instance, the large scale orbit of the stream is typically governed by \emph{e.g.} the ambient gravitational potential, as well as the initial conditions of the cluster. On the other hand, the smaller-scale features (gaps, spurs etc.) are more likely to be impacted by the dynamical evolution history, tidal stripping, or interactions with perturbers. The aim is to analyse both of these classes of signal simultaneously, and as such we should choose a network architecture accordingly. In particular, with this observation, it is simple to see that applying \emph{e.g.} a standard convolutional network which applies the same kernel to each part of the image identically would be a poor choice and unlikely to be able to simultaneously extract the small scale and large scale information. With this in mind, we choose to use the well-known \texttt{unet} architecture~\cite{Ronneberger:2015aaa}, which is well suited for image analysis and segmentation. It is designed to simultaneously analyse the image at a larger scale, before performing follow up analysis on each identified segment and then combining the results.

There is another part of the inference network (a full description and network diagram can be found in Appendix~\ref{app:network}) which performs the ratio estimation. Schematically, one can understand the overall structure as first performing some data compression through the \texttt{unet} and a small linear network to extract an optimal set of summary statistics. Then, these summary statistics (which are \emph{automatically} learned and optimised during the training), are passed to the default ratio estimator implemented in \texttt{swyft} along with the model parameters $\theta$. All the details regarding the implementation can be found in the \texttt{albatross} library. In terms of specificity, we expect the network to be broadly applicable to the analysis of any stream model or observation, since it only assumes that the signal has structure on various scales.

\vspace*{8pt}
\noindent \emph{Prior Choices.} The prior choices for all the parameters of interest are shown in Tab.~\ref{tab:parameters}. They are chosen to either represent our knowledge about the physics from current astrophysical observations or simulation results (\emph{e.g.} the mass loss parameter $\alpha$), or to be maximally uninformative. An example of the latter case are the cluster position and velocity priors which are chosen in the first instance to span the full observational window.

\vspace*{8pt}
\noindent \emph{TMNRE Hyperparameters.} There are a number of hyperparameters that need to be set when one runs the TMNRE algorithm. Broadly these can be categorised as either parameters that control the network training process, or parameters specific to the TMNRE algorithm. For the inference and analysis detailed in this work, the particular choices can be found in Tab.~\ref{tab:sbi_tmnre}, as well as in the example configuration files supplied with \texttt{albatross}. Briefly, the training parameters describe how long to train the network for (min./max. training epochs), how many epochs to wait before the validation loss should decrease again (early stopping patience)\footnote{During the training, we track both the current loss on the training data set, as well as the loss evaluated on some separate validation set. Looking for good performance on the validation data set is typically a good strategy to avoid overfitting, and therefore we use it as a metric to indicate whether we are starting to overfit to the training data. The early stopping criterion waits for a specified number of passes through the training data (or epochs), over which the validation loss has not decreased before terminating the training. It then re-initialises the network parameters to the state where the minimum validation loss was observed.}, the split between training and validation data (Train : Validation ratio), and the batch sizes shown to the network during training (training/validation batch size). The TMNRE settings consist of the minimum number of rounds (number of rounds), the schedule for the number of simulations per round (simulation schedule)\footnote{It is typically the case that in the early rounds, only a small number of parameters are meaningfully constrained, and so it is more efficient to have a more reduced simulation batch, truncate, and then re-simulate again. In the last rounds, however, to achieve the correct level of statistical precision, significantly more training data is required.}, and the threshold for truncation ($\epsilon$). Finally, we have the ``noise shuffling" setting, which breaks down the data $x$ into the stream and background components. In a given batch it then randomly permutes the background elements, essentially showing the network a brand new example (with the same signal component) every epoch at zero simulation cost. We found this to be an extremely effective way of reducing the possibility of overfitting, especially in the early rounds where we have small simulation batches\footnote{As an aside, we also explicitly tested that resampling the stripping times and regenerating the ``same" stream (at least statistically) also lead to improvements in the smoothness of the training and validation losses, but importantly did not affect the precision of the posteriors. This approach was especially effective for small simulation batches.}. Indeed, this strategy should be applicable to any additive noise model, see for example its application to gravitational wave data analysis~\cite{Bhardwaj:2023xph}.

\begin{table}[t]
\centering
\resizebox{\linewidth}{!}{
\begin{tabular}{@{}ll@{}}
\hline
\textbf{TMNRE Setting}          & \textbf{Value} \\ \hline
Number of rounds                & 7${}^\star$ \\
Simulation schedule             & 30k, 30k, 30k, 30k, 30k, 60k, 150k \\
Bounds threshold $\epsilon$     & $10^{-5}$ \\ 
Noise shuffling                 & \texttt{True} \\ 
Min./Max. training epochs       & 0/50 \\ 
Early stopping patience         & 20 \\ 
Initial learning rate           & $5 \times 10^{-4}$ \\
Training/Validation batch size  & 64/64 \\ 
Train : Validation ratio        & 0.9 : 0.1 \\ \hline
\end{tabular}}
\caption{Choices for the hyperparameters and settings for the TMNRE algorithm in this work, as described in Sec.~\ref{sec:sbi}.\\
{\footnotesize ${}^\star$This is the minimum number of rounds, if the algorithm has not converged, we continue rounds of inference until the truncation procedure terminates.}}
\label{tab:sbi_tmnre}
\end{table}

\begin{figure*}[t]
    \centering
    \includegraphics[width=0.7\linewidth,trim={0.1cm 0.1cm 0.1cm 0.1cm},clip]{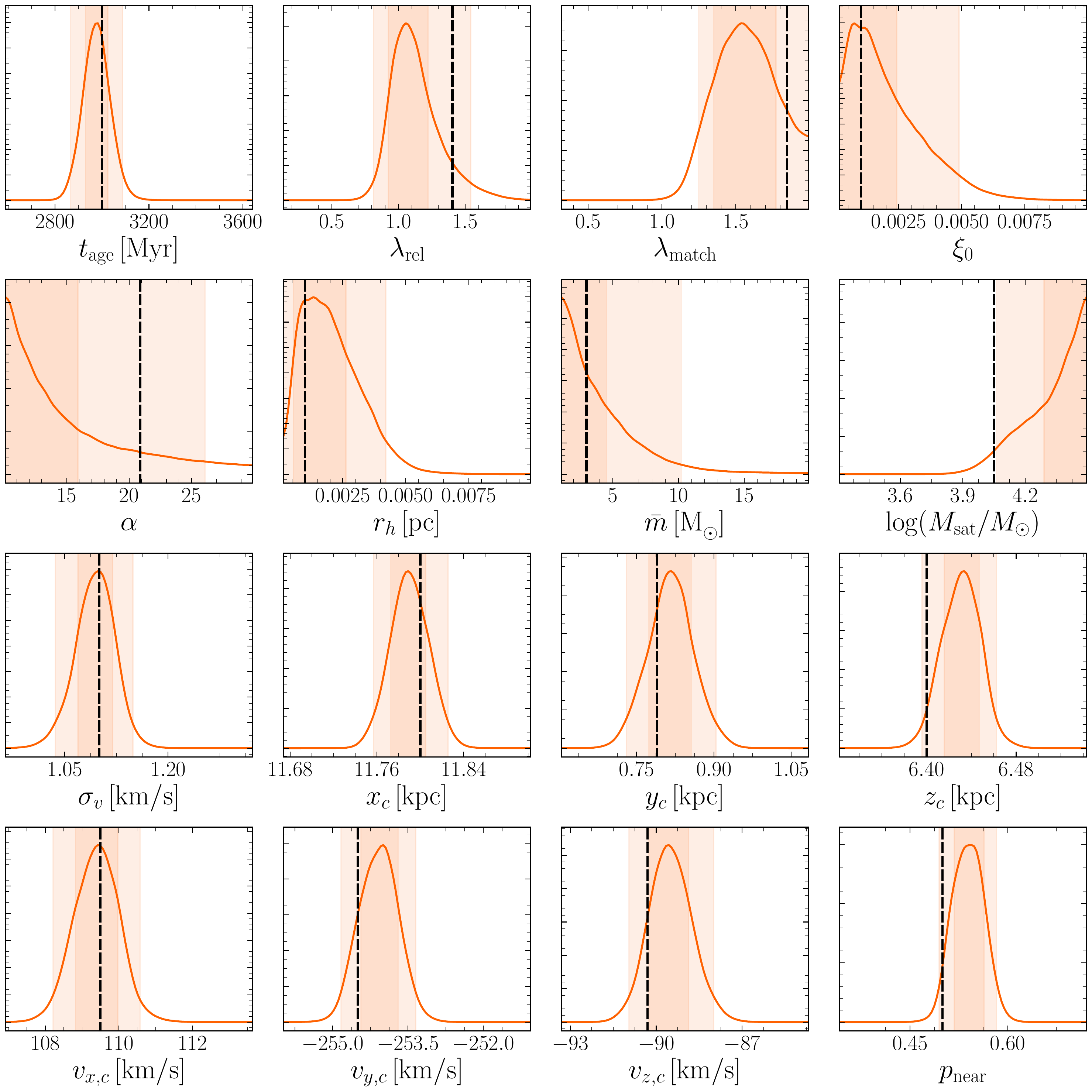}
    \caption{Full set of 1d marginal posteriors (orange curves) for all parameters in the \texttt{sstrax} stream model described in Sec.~\ref{sec:stellarstreams} applied to the mock observation in Fig.~\ref{fig:observation}. The $1\sigma$, $2\sigma$, and $3\sigma$ contours are overlaid behind. Finally, the black vertical lines indicate the true injected parameters from Tab.~\ref{tab:parameters}.}
    \label{fig:results}
\end{figure*}

In this section we have discussed the broad field of simulation-based inference and a specific algorithm, known as Truncated Marginal Neural Ratio Estimation (TMNRE) that we have used to build our data analysis pipeline. We argued that the targeted and marginal-focused approach could be a key advantage for stellar stream analysis, including the resulting simulation efficiency, statistical robustness, and the opportunities for increased model complexity. Finally, we discussed some of the design choices that need to be made in order to successfully apply TMNRE to a given problem. In the next section, we will present a case study for a mock stream to illustrate the application of our modelling and analysis strategy. 

\section{Results: GD1-like Case Study}\label{sec:results}

\noindent Now that we have set up the framework of simulation-based inference, and specifically described the application of the algorithm to the analysis of stellar streams, we can present a case study to highlight its functionality. In this section, we will illustrate the full analysis and validation of a mock observation that is generated using our stellar streams modelling code \texttt{sstrax}. This is in order to have full control over the reconstruction of the parameters, as well as the physics input to the model. Of course, the longer term goal is to analyse current and future state-of-the-art spectroscopic and photometric surveys, such as \emph{Gaia} and the Vera Rubin observatory~\cite{Gaia:2018aaa,Gaia:2021aaa,LSSTScience:2009jmu,Bechtol:2019acd}.

\begin{figure*}[t]
    \centering
    \includegraphics[width=0.9\linewidth,trim={0.1cm 0.1cm 0.1cm 0.1cm},clip]{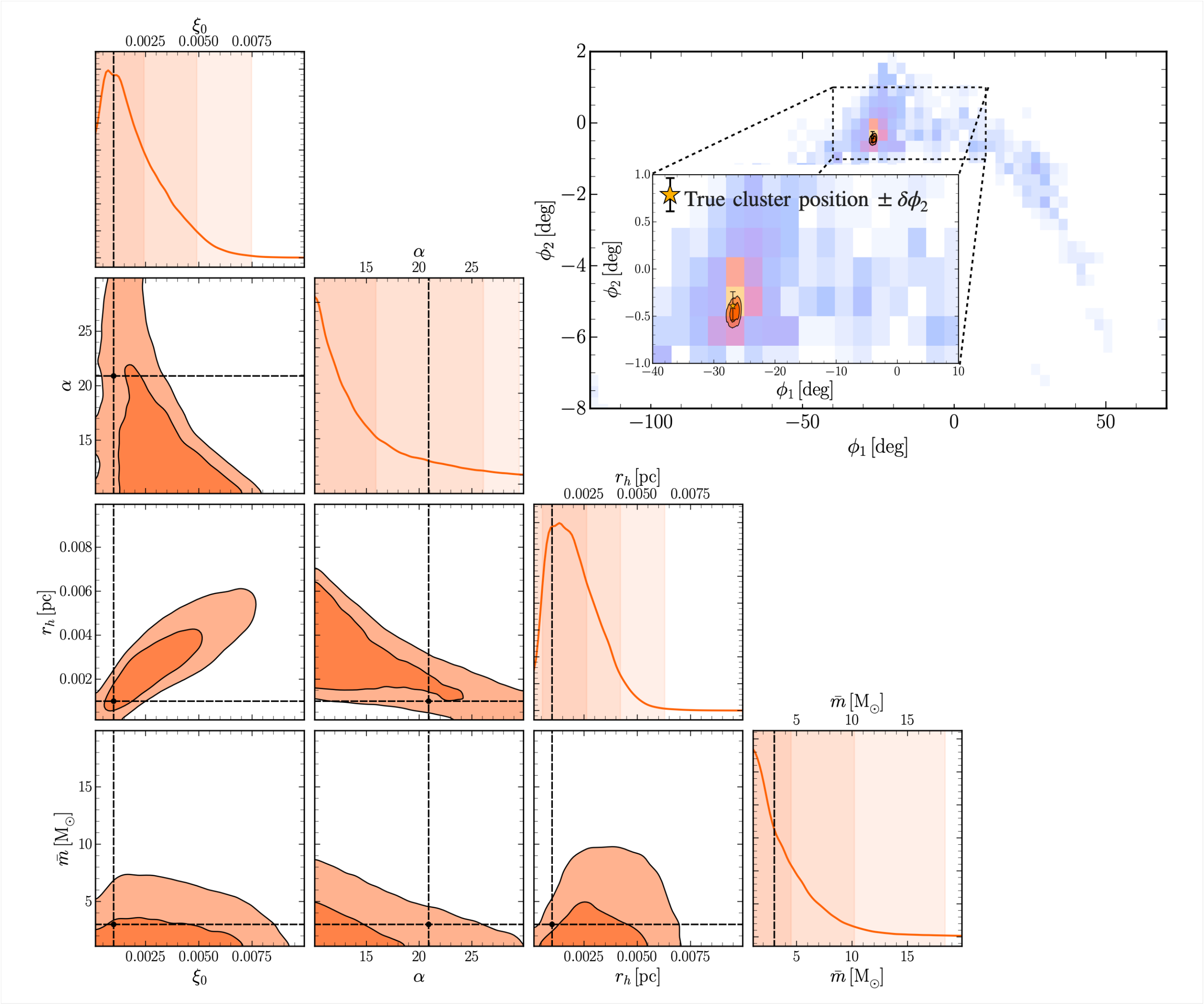}
    \caption{\emph{Corner plot:} Follow up analysis on the mass loss model given the 1d marginals in Fig.~\ref{fig:results}. The orange contours show the trained 2d posteriors on the parameters relevant to the mass loss model ($\xi_0$, $\alpha$, $r_h$, $\bar{m}$). \emph{Upper right panel and inset:} Derived marginal posteriors (orange contours) in the $(\phi_1, \phi_2)$-plane on the final cluster position overlaid on top of the mock target observation. In addition, we highlight the observational errors (black error bars) and the true value (yellow star) to be reconstructed.}
    \label{fig:corner_plot}
\end{figure*}

\subsection{Case Study Description}\label{sec:results_casestudy}

\noindent Perhaps the most well-studied and well-observed Milky Way stellar stream is the GD1 stream~\cite{Price-Whelan:2018aaa,Grillmair:2006bd,Eyre:2009hg}. It was first identified in the SDSS catalog in the early 2000s~\cite{Grillmair:2006bd,Eyre:2009hg}, but recently it has been observed by \emph{e.g.} \emph{Gaia} in significantly more detail~\cite{Price-Whelan:2018aaa,Gaia:2018aaa,Gaia:2021aaa}. Indeed, observations are currently at the level where individual substructures (\emph{e.g.} the so-called ``gaps" and ``spur") are reasonably well resolved~\cite{Price-Whelan:2018aaa,deBoer:2018aaa}. In some sense, the purpose of developing our analysis method is to take full advantage of these improvements and perform inference on streams like GD1 in as realistic as possible simulation framework.

To illustrate and test our method, we construct a case study to closely resemble the sky location and structure of the GD1 stream. To do so, we choose a stream closely aligned with the $\phi_2 = 0\,\mathrm{deg}$ plane in the GD1 specific co-ordinate system defined in Sec.~\ref{sec:stellarstreams}. We centre the location of the cluster (remnant) at $\phi_1 \sim -25 \, \mathrm{deg}$, and choose the age of the stream such that it extends across a significant portion of the sky, as in the case of the real GD1 observation~\cite{Grillmair:2006bd,Eyre:2009hg,Price-Whelan:2018aaa}. Similarly, the dominant part of the stream is located around $6-10\,\mathrm{kpc}$ away from the galactic centre. The full set of parameter values that we choose for the mock observation are shown in Tab.~\ref{tab:parameters}, along with the priors for the subsequent Bayesian inference. The mock observation that these parameters generate, and the focus of the analysis below is shown in Fig.~\ref{fig:observation}. We do note, however, that whilst the mock observation we present here has general features that represent a GD1-like stream, it is important to acknowledge that some aspects of the modelling could be improved in this regard. One important example is the presence of a clear remnant at the cluster centre, which is not present in the real GD1 data. Concretely, this has the effect that the reconstruction of the cluster position in our analysis may be overly optimistic compared to a more realistic case.

\subsection{Parameter Estimation with TMNRE}\label{sec:results_pe}

\begin{figure*}[t]
    \centering
    \includegraphics[width=\linewidth,trim={0.1cm 0.1cm 0.1cm 0.1cm},clip]{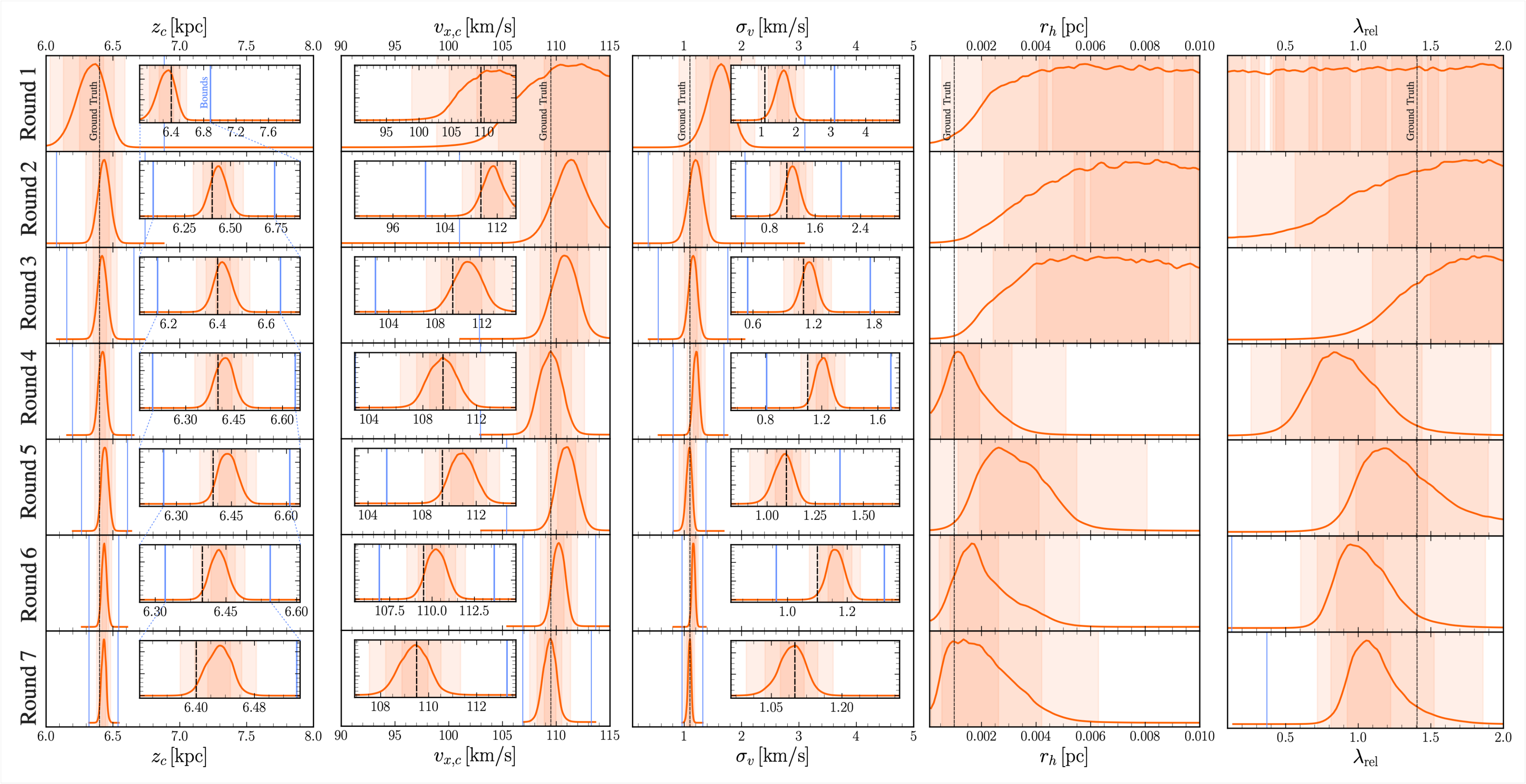}
    \caption{Examples of the truncation procedure in TMNRE applied to five of the model parameters. From \emph{left} to \emph{right}, we illustrate the evolution of the posterior estimates for $z_c$, $v_{x,c}$, $\sigma_v$, $r_h$, and $\lambda_\mathrm{rel}$. From \emph{top} to \emph{bottom} we show the development over the number of rounds of the TMNRE algorithm. The \emph{insets} zoom in on the bounded region (blue vertical lines) to highlight the coverage of the true value (vertical black dotted line).}
    \label{fig:truncation}
\end{figure*}

\noindent We carry out parameter estimation using the priors for the model parameters indicated in Tab.~\ref{tab:parameters}, the observational model described in Tab.~\ref{tab:obs_parameters}, and the TMNRE algorithm settings given in Tab.~\ref{tab:sbi_tmnre}. The key results for this section are given in Figs.~\ref{fig:results},~\ref{fig:corner_plot}, and~\ref{fig:truncation}.

\vspace*{8pt}
\noindent\emph{Overview.} There are a few levels at which to discuss our results from applying the TMNRE algorithm described in Sec.~\ref{sec:sbi} to the mock GD1-like observation in Fig.~\ref{fig:observation}. The first is simply in the context of robust and faithful inference -- in Fig.~\ref{fig:results}, we show the full set of converged 1d-posteriors for all parameters in the model. We see that we reconstruct the true value in all cases either via a direct measurement (\emph{e.g.} the final position or velocity of the cluster) or as some clear upper or lower bound (\emph{e.g.} the mass loss parameter $\alpha$). Importantly, we can reconstruct with very high precision the age of the stream ($t_\mathrm{age}$), the velocity dispersion of the cluster ($\sigma_v$), the current cluster position and velocity ($[x_c, y_c, z_c], [v_{x,c}, v_{y,c}, v_{z,c}]$), and the relative asymmetry in the tidal stripping ($p_\mathrm{near}$). This sort of constraint is easy to motivate physically via \emph{e.g.} the length and width of the stream, which is strongly affected by the age and velocity dispersion, as well as the stream's spatial location and orientation which is controlled by the relative cluster position and velocity\footnote{Indirectly, the exact orientation will be affected, of course, by the gravitational potential of the Milky Way, see \emph{e.g.}~\cite{Koposov:2009hn,Bowden:2015jva,S5:2021edi,Sanderson:2014apa,Erkal:2019aaa,Gibbons:2014ewa,Bovy:2016chl,Bonaca:2014qia}, which we have fixed in this analysis. It is easy to generalise to a case where the potential is allowed to freely vary with an analysis pipeline that would remain identical.}.

\vspace*{8pt}
\noindent\emph{Degeneracies.} Of course, not every parameter is measured with high precision, such as the parameters ($\xi_0$, $\alpha$, $r_h$, $\bar{m}$) that control the mass loss rate of the cluster as it orbits the Milky Way. Whilst we can set relevant upper or lower bounds on these parameters\footnote{Which again we might be able to motivate physically -- for example, it makes sense that we can set a lower bound on the total mass of the cluster just by having some count of the total number of observed stream stars and multiplying by the average stellar mass.}, it is interesting to explore the degeneracies between these parameters also. This is where we can use the flexibility of the TMNRE algorithm to efficiently estimate posteriors of choice. Specifically, we only need to estimate the relevant 2d posteriors for exploring the degeneracy structure of the mass loss model. To do so, we train additional ratio estimators $r(x; \{\theta_i, \theta_j\})$ with $\theta_i, \theta_j \in (\xi_0, \alpha, r_h, \bar{m})$. The results for these 2d-posteriors, along with the corresponding 1d-posteriors are shown in Fig.~\ref{fig:corner_plot}. We see some clear degeneracies highlighted in the parameter inference such as those between $\xi_0$ and $r_h$. Again, whilst we do not investigate these degeneracies in detail, this could be expected from \emph{e.g.} the scaling of $\mathrm{d}M_c/\mathrm{d}t$ in Eq.~\eqref{eq:stellarstreams_massloss}.

\vspace*{8pt}
\noindent\emph{Precision.} As discussed in Sec.~\ref{sec:sbi}, simulation-based inference has now been used extensively in other fields, and has been shown to qualitatively and quantitatively reproduce known results and results obtained using traditional methods. In the present case, a full comparison to \emph{e.g.} a traditional method such as MCMC is challenging because we only have a forward simulator for the observational and noise models. This is another way to say that we do not have an explicit form of the data likelihood. Of course, this is a key strength of the class of simulation-based methods, since it allows for arbitrarily complex data simulators, which can account for complicated aspects of detection and selection in a statistically meaningful way. On the other hand, this means we should consider additional ways to test our results. A simple qualitative test we can perform is to compare the precision (and accuracy) with which we reconstruct the cluster position to the intrinsic observational errors on the stellar positions. To test this, we construct the 3d-joint posterior $p(x_c, y_c, z_c | x_0)$ by training a 3d ratio estimator $r(x; \{x_c, y_c, z_c\})$ on the final round of simulations. From this joint posterior, we can generate posterior samples in the $(\phi_1, \phi_2)$ parameter space\footnote{Note that, of course, it would have been statistically incorrect to generate these from the individual marginal posteriors on the cluster positions, even though they are well measured.} for the current position of the cluster that are distributed as $\phi_1, \phi_2 \sim p(\phi_1, \phi_2 | x_0)$. These are shown in the top right panel (and inset) of Fig.~\ref{fig:corner_plot} along with the observational model errors on the positions of the stars $\delta\phi_1, \delta\phi_2$. We see that we are able to reconstruct the cluster position to a good degree of accuracy and precision.

\vspace*{8pt}
\noindent\emph{Simulation Efficiency.} One of the key arguments we made for using TMNRE was the fact that it gave us the ability to use high fidelity simulators. This is both from a statistical perspective in the sense that we can perform Bayesian inference without explicit likelihoods, but also from the scalability point of view. Indeed, one of the main obstacles for a full analysis of stellar streams is that fact that performing enough simulations to do inference on a large number of parameters is typically infeasible. This is where the marginal and targeted aspects of TMNRE are relevant, as well as the acceleration of the simulator. To be more specific, in the case study described above, we required a total of only $350$k simulations to perform inference on all 16 parameters simultaneously. Crucially, this simulation budget was split across a total of 7 rounds, as illustrated in Tab.~\ref{tab:sbi_tmnre}. In between the rounds, the truncation procedure described in Sec.~\ref{sec:sbi} was applied, which ensures that we are targeting the specific observation of interest, and that the variance in the training data is significantly reduced compared to the previous round. This is very important for simulation efficiency, and results in much higher quality inference results on targeted observations compared to \emph{e.g.} the case where a fixed simulation budget is used in a single round\footnote{Of course, if the goal is to perform some sort of amortised inference across all possible observations, then one should use this hypothetical simulation budget differently. For an example in the context of gravitational waves, see \emph{e.g.}~\cite{Bhardwaj:2023xph}.}. This truncation process is highlighted in Fig.~\ref{fig:truncation}, where we see how the different classes of parameter respond to the truncation process. For example, the first three columns of parameters (one component of the position and velocity, and the velocity dispersion of the stream) are extremely well constrained once the algorithm converges. On the other hand, the last two panels show parameters that are only broadly reconstructed ($r_h$ and $\lambda_\mathrm{rel}$). For this second class of parameter, however, we see that in the initial rounds, the marginal posterior estimates of \emph{e.g.} $p(r_h | x_0)$ are quite poor\footnote{In fact, it is a good example of where we should be careful not to interpret these early-round ratio estimators as strict posteriors, since the algorithm has not converged.}. As the rounds evolve and the well-measured parameters are better constrained, subsequently reducing the training data variance, the posterior estimates on the poorly reconstructed parameters significantly improve. This is a general feature of TMNRE, where convergence and truncation in one set of parameters leads to marked improvements in the inference of other model parameters, even if they themselves are not well measured.

\begin{figure}[t]
    \centering
    \includegraphics[width=0.8\linewidth,trim={0.1cm 0.1cm 0.1cm 0.1cm},clip]{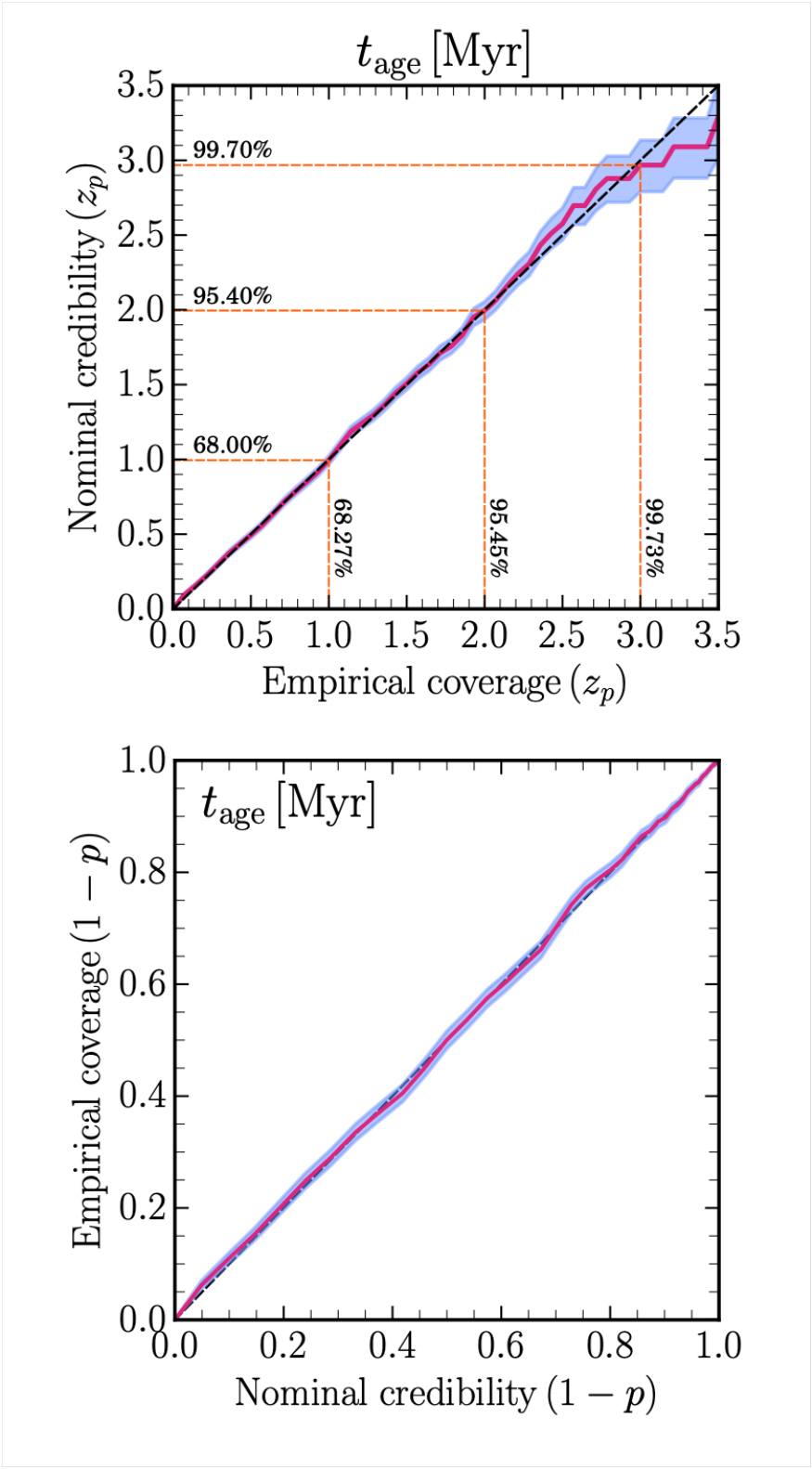}
    \caption{Example of the coverage tests applied in this work for the age of the stream $t_\mathrm{age}$. \emph{Top panel:} Empirical (observed) against the nominal or expected coverage. \emph{Bottom panel:} The same information as the top panel but plotted in terms of the corresponding $p$-values. The red lines indicate the actual coverage results, whilst the blue contours represent the 1$\sigma$ confidence interval on this estimate.}
    \label{fig:tage_coverage}
\end{figure}

In terms of actual run time, we performed this analysis on a 72 CPU core cluster node, with a single \texttt{NVIDIA A100} GPU to train the ratio estimators. The total run time for the analysis was around 19 hours, of which approximately 90\% was simulation time. Note that this can therefore be improved immediately by either \emph{(a)} further speeding up the simulator, or \emph{(b)} having access to more CPU cores where the simulations can be further parallelised.

\subsection{Consistency and Validation Tests}\label{sec:results_validation}

\noindent The posterior sanity checks and explicit evidence for excellent reconstruction of the true values for the parameters in our case study are an important step towards developing and testing our analysis pipeline. On the other hand, given that our goal is to target data analysis challenges where there are no traditional methods available -- either because they scale too poorly with the number of parameters, or because they have an analytically intractable data likelihood -- we need to develop additional consistency checks to validate our results. This is very much an active field of research in simulation-based inference, and a set of established methods now exist~\cite{Hermans:2021aaa,Lueckmann:2021aaa}.

The most common, and the one that we will present here, are known as coverage tests~\cite{Hermans:2021aaa}. We will focus on expected coverage tests of our inference pipeline, which make precise the idea of variations in posterior estimates over various observational or statistical fluctuations. In particular, expected coverage tests ask the following question: \emph{how often does the $x$\% credible interval contain the true value, averaged over observations generated from the joint distribution $x, \theta \sim p(x, \theta)$?} By definition, a well calibrated posterior distribution will contain the true value inside the $x$\% credible internal $x$\% of the time. As such, to carry out this test, we can generate a set of simulations\footnote{Here we generate 1000 new simulations.} from the truncated prior in the final round of inference (so that we test the most relevant region of parameter space), and then perform inference on each simulation using our trained final round ratio estimators. For each confidence level $x$\% $\in [0, 1]$, we can then count how many simulations have inference results that contain the corresponding true value within this confidence interval. A posterior will pass this test if this results in an approximately diagonal line in the \emph{expected} versus \emph{empirical} coverage plane. Importantly, since the inference must be done individually for each mock observation, it is typically infeasible to perform this sort of coverage test with fully sequential methods (including \emph{e.g.} MCMC or nested sampling), especially in scenarios with high simulation cost such as stellar streams. Finally, one should note that this coverage test is diagnostic in the sense that a failure indicates a poorly calibrated posterior estimate, but success does not guarantee that the correct posterior has been found.

We provide the coverage test results for all 16 parameters in the Appendix (see Figs.~\ref{fig:zz} and~\ref{fig:pp}), but also give a specific example for the age of the stream $t_\mathrm{age}$ in Fig.~\ref{fig:tage_coverage} opposite. We see that in all cases we achieve good coverage results, which can easily be improved further by allocating a slightly larger simulation budget. This coverage test diagnostic will remain applicable irrespective of the forward model or parameter choices, and is one of the key metrics for being able to validate simulation-based inference methods.

\section{Conclusions and Outlook}\label{sec:conclusions}

\noindent In this work, we have presented the development and application of a brand new simulation-based inference data analysis pipeline for modelling (see Sec.~\ref{sec:stellarstreams}) and analysing stellar streams (see Secs.~\ref{sec:sbi} and~\ref{sec:results}). In this last section, we present our key conclusions and provide some outlook as to the classes of analysis challenge we can now attempt to tackle, as well as the steps that would be required to achieve them. The key contributions in the work are as follows:
\begin{itemize}[leftmargin=*]
    \item \emph{Scalable simulation-based inference pipeline.} We have developed a brand new simulation-based inference algorithm to analyse stellar streams in the Milky Way (see Sec.~\ref{sec:sbi} for a discussion on the application of simulation-based methods to stellar streams). In particular, we have implemented the TMNRE algorithm~\cite{Miller:2022shs} with the aim to develop a scalable inference method. The motivation for choosing this algorithm for the analysis of stellar streams is mainly due to simulation efficiency that results from targeting individual observations and focusing on marginals. We showed in Sec.~\ref{sec:results} that we were able to perform inference on all 16 parameters of our model with only 350k simulations. This sort of performance is the key argument for the ability of our approach to analyse streams with far more realistic forward models.
    \item \emph{Robust and flexible method.} Another important aspect of the analysis methodology developed here is its flexibility and robustness to changes in observation strategy or simulation model. By definition, our approach is simulation-based and therefore has the advantages that it does not \emph{(i)} assume any explicit likelihood for \emph{e.g.} the observational model, only the existence of a forward simulator, and \emph{(ii)} assume any particular form of the data output. On the latter point, whilst we have developed the algorithm alongside our modelling code \texttt{sstrax}, the analysis pipeline would remain \emph{identical} for any simulator. This is the crucial aspect that will allow our method to be used for making direct comparisons between different stream simulation strategies and observational models.
    \item \emph{Public analysis code.} We have built our analysis method on top of the \texttt{swyft} software which is a \texttt{pytorch}-based implementation of TMNRE~\cite{Miller:2022shs}. Specifically, we have publicly released the \href{https://github.com/undark-lab/albatross}{\texttt{albatross}} code that is currently coupled to the \texttt{sstrax} modelling code by default. The \texttt{albatross} code is highly modular and can in principle be coupled to any forward model, for example \texttt{galpy}~\cite{Bovy:2015aaa}, without any change in the analysis methodology. This will eventually allow for direct comparisons in the inference between different modelling strategies.
    \item \emph{Public modelling code.} As mentioned above, one of the key motivations for developing the \texttt{albatross} implementation of the TMNRE algorithm was to create a framework that allows for robust, scalable inference on complex models. In the same vein, we developed a new modelling code \href{https://github.com/undark-lab/sstrax}{\texttt{sstrax}} that is accelerated through the \texttt{jax} programming paradigm~\cite{jax:2018aaa}. This allows for fast (around a second per simulation) and realistic forward modelling of streams. We have designed the code to be readily extendable to include any physical effects such as subhalo impacts, varying gravitational potentials, higher fidelity tidal disruption models etc. As above, regardless of the modelling choices, the inference pipeline will crucially remain identical.
\end{itemize}

\subsection{Outlook}\label{sec:conclusions_outlook}

\noindent We argued in the introduction that stellar streams are an exciting probe of galactic and dark matter physics. This is particularly true as the quality of observations continues to significantly improve in the eras of \emph{Gaia} and the Vera Rubin observatory~\cite{Gaia:2018aaa,Gaia:2021aaa,LSSTScience:2009jmu,Bechtol:2019acd}. Taking full advantage of this data is challenging, however, both in terms of robust statistical analysis and the complexity of simulations required. Ultimately, if we are interested in using stellar streams to analyse scenarios such as the origin and statistics of substructure in the stream~\cite{Banik:2018pal,Banik:2018pjp,Banik:2019smi,Banik:2019cza}, or the impact of a large population of low mass subhalos on streams in the Milky Way~\cite{Erkal:2014tda,Erkal:2015kqa,Bonaca:2018fek}, we will have to overcome these hurdles. This is the context we had in mind when developing \texttt{albatross} and \texttt{sstrax}. The aim was to develop a scalable, simulation-efficient framework that did not make any assumption about the complexity of the forward simulator. This is exactly the sort of task that simulation-based inference methods were developed to address. In terms of specific outlook, we believe there are a number of interesting avenues to pursue given the capabilities developed here.

On the analysis side, there are a number of interesting claims in the literature about the origin and characterisation of the gaps and features in streams such as GD1~\cite{Grillmair:2006bd,Eyre:2009hg,deBoer:2018aaa,Price-Whelan:2018aaa}. More specifically, it would be an extremely important result to classify \emph{e.g.} the gap in GD1 as being due to a compact object or subhalo collision~\cite{Price-Whelan:2018aaa,deBoer:2018aaa,Carlberg:2013gxa}. To obtain a definitive answer, however, one needs to show that the features cannot (at least to some degree of statistical certainty) arise by chance as a result of some stochasticity in the tidal stripping process, selection effects at the level of stream detection, or as a result of a more complex model of the Milky Way potential including known substructure such as dwarf galaxies or globular clusters~\cite{Doke:2022jro,Dillamore:2022aaa,Amorisco:2016evb}. Similarly, it would be interesting to provide a conclusive answer as to the relative shape and size of the Milky Way gravitational potential from an analysis of individual or multiple streams~\cite{S5:2021edi}. The key advantage of the framework we have put forward here is that one can (and should) ask all of these questions simultaneously. This is a more precise version of the statement in the introduction where we argued that we would ideally like to analyse the large- and small-scale structures present in stellar streams at the same time. On a more cautionary note, in order to move towards analysing real data in its full complexity with this class of simulation-based inference methods, it will be important to characterise and quantify the sensitivity of the inference method to perturbations or misspecifications in the forward model. This is particularly relevant in the case of stellar streams where the physics is highly complex, and it is unlikely to be possible for a simulation model to be developed that is simultaneously fully self-consistent \emph{and} fast enough for parameter inference. One step towards this goal could include analysing mock streams generated from N-body simulations (see \textit{e.g.}~\cite{Varghese:2011aaa}) with identifiable parameters that match those in the model (such as the cluster velocity dispersion, analytic potential, or the age of the stream). One should bear in mind, however, that this is not necessarily an issue directly with simulation-based inference, but also affects traditional approaches such as MCMC if \textit{e.g.} the modelling or data likelihood is miscalibrated.

Of course, to achieve these analysis goals, we must also make progress on modelling. Having a flexible analysis and simulation pipeline that does not assume \emph{e.g.} symmetry in the Milky Way potential, or uniform stripping times in the evolution of the cluster motivates us to focus on improving the realism of each aspect. In particular, there are a number of key developments that would place the analysis questions above on a much more solid footing and allow us to analyse real data with confidence. Firstly, we should focus attention on the observational model -- in this work we constructed a very simple framework to describe the detection and measurement of Milky Way streams. In reality, however, data such as that from \emph{Gaia} is significantly more complicated~\cite{Gaia:2018aaa,Gaia:2021aaa}, accounting for \emph{e.g.} position dependent errors, selection effects based upon proper motions and metallicities, and spatially varying background densities~\cite{Gaia:2018aaa,Gaia:2021aaa}. Realistic modelling of this will be particularly relevant for robustly studying \emph{e.g.} small-scale features in streams. Secondly, the dynamics of tidal disruption and the release of stars from the cluster is vital for generating realistic density perturbations along the stream track. Again, since this could be an interesting observable for studying \emph{e.g.} the collective implications of a population of small perturbers~\cite{Banik:2018pjp,Bonaca:2018fek,Banik:2018pal,Delos:2021ouc}, or the internal properties of globular clusters~\cite{Gialluca:2020tno}, development of the model realism will inevitably lead to more informative inference results. Thirdly, we know that on the sort of timescales relevant to stellar streams, the Milky Way and its potential are dynamical, both in terms of its global structure, as well as the large amount of substructure in the form of dwarf galaxies, other clusters, or gas clouds~\cite{Dillamore:2022aaa,Amorisco:2016evb,Doke:2022jro}. It would be interesting to take input from \emph{e.g.} N-body simulations of Milky Way formation and trace the evolution of streams in such a dynamical potential. As we argued above, this could be done without any change in the analysis pipeline. 

In summary, the development of a scalable and flexible simulation-based inference approach to analysing stellar streams can allow us to answer important questions about the evolution of, and substructure in our own galaxy. Aided by high quality observations by the latest surveys~\cite{Gaia:2018aaa,Gaia:2021aaa,LSSTScience:2009jmu,Bechtol:2019acd}, we can use this to start asking concrete questions regarding the nature of dark matter, the evolution and structure of the Milky Way, or the dynamics of dwarf galaxies and globular clusters. To achieve this will require development from the perspective of modelling stream dynamics and survey observations. However, having a robust simulation efficient inference strategy is strong motivation for starting to move further towards this ambitious goal.

\vspace*{8pt}
\section*{Acknowledgements}
\noindent JA is supported through the research program ``The Hidden Universe of Weakly Interacting Particles" with project number 680.92.18.03 (NWO Vrije Programma), which is partly financed by the Nederlandse Organisatie voor Wetenschappelijk Onderzoek (Dutch Research Council). CW and MG are supported by a project that has received funding from the European Research Council (ERC) under the European Union's Horizon 2020 research and innovation programme (Grant agreement No. 864035). We are extremely grateful to Noemi Anau Montel, Gianfranco Bertone and Sam Witte for helpful discussions regarding this project. The main analysis for this work was carried out on the Lisa and Snellius Computing Clusters at SURFsara.

\vspace*{8pt}
\section*{Data Availability}
\noindent The code for generating the data underlying this manuscript is available in the \texttt{sstrax} library (at \href{https://github.com/undark-lab/sstrax}{this link}) and the \texttt{swyft}-based inference library \texttt{albatross} (available \href{https://github.com/undark-lab/albatross}{here}).

\bibliography{biblio}
\newpage
\appendix

\section{Co-ordinate transformations in \texttt{sstrax}}\label{app:sstrax}

\noindent Here we detail the co-ordinate transformations we use in \texttt{sstrax} to move from the cartesian simulation frame $X_{\mathrm{halo}} \equiv (x, y, z)$ to the GD1 co-ordinates $(r, \phi_1, \phi_2)$. This is implemented in the \texttt{projection.py} module, and is explicitly given by the following set of relations,
\begin{equation}
    X_\mathrm{halo} \equiv (x, y, z),
\end{equation}
Then, in a frame centred at the sun with $x_\mathrm{sun} = 8\,\mathrm{kpc}$,
\begin{equation}
    X_\mathrm{sun} \equiv (\tilde{x}, \tilde{y}, \tilde{z}) = (x_\mathrm{sun} - x, y, z).
\end{equation}
We can convert to galactic co-ordinates $X_\mathrm{gal} \equiv (r, b, l)$ via,
\begin{align}
    r &= \sqrt{\tilde{x}^2 + \tilde{y}^2 + \tilde{z}^2}, \\
    b &= \arcsin(\tilde{y}/r), \\
    l &= \arctan(\tilde{y}/\tilde{x}).
\end{align}
Then, we can rotate to equatorial co-ordinates $X_\mathrm{equat} \equiv (r, \alpha, \delta)$ through,
\begin{align}
    \alpha &= \tan^{-1} \left(\frac{\cos b \sin (l_\mathrm{NGP} - l)}{\cos \delta_{\mathrm{NGP}} \sin b - \sin \delta_{\mathrm{NGP}} \cos b \cos(l_\mathrm{NGP} - l)}\right) \nonumber \\
    & \qquad \qquad \qquad \qquad + \alpha_\mathrm{NGP} \\
    \delta &= \arcsin \left( \sin \delta_{\mathrm{NGP}} \sin b + \cos \delta_{\mathrm{NGP}} \cos b \cos(l_\mathrm{NGP} - l)\right),
\end{align}
with $\delta_\mathrm{NGP} = 27.12825118085622 \, \mathrm{deg}$, $l_\mathrm{NGP} = 122.9319185680026 \, \mathrm{deg}$, and $\alpha_\mathrm{NGP} = 192.85948 \, \mathrm{deg}$. After this, we can rotate to a cartesian co-ordinate frame aligned with the stream $X_\mathrm{gd1,cart} \equiv (x_g, y_g, z_g)$ with,
\begin{align}
    \begin{pmatrix} x_g \\ y_g \\ z_g \end{pmatrix} &= \begin{bmatrix}
-0.4776303088 & -0.1738432154 & 0.8611897727 \\
0.510844589 & -0.8524449229 & 0.111245042 \\
0.7147776536 & 0.4930681392 & 0.4959603976 
\end{bmatrix} \nonumber \\ 
& \qquad \qquad \times \begin{pmatrix} r \cos \alpha \cos \delta \\ r \cos \alpha \sin \delta \\ r \cos \delta \end{pmatrix},
\end{align}
taken from Ref.~\cite{Koposov:2009hn}. Finally, we can construct our GD1 co-ordinates $X_\mathrm{gd1} \equiv (r, \phi_1, \phi_2)$ via,
\begin{align}
    \phi_1 = \arctan(y_g/x_g), \,\, \phi_2 = \arcsin(z_g / r).
\end{align}
The final step in the co-ordinate transformations is to construct the velocity in a different co-ordinate frame given the velocity in the simulation frame. For this, we take advantage of the auto-differentiation capability of \texttt{jax} and numerically compute the Jacobian $\mathcal{J}_\mathrm{ij} \equiv \partial X^i_\mathrm{gd1} / \partial X^j_\mathrm{halo}$. The velocity in the GD1 co-ordiante frame $V_\mathrm{gd1} \equiv (v_\mathrm{rad}, \dot{\phi}_1, \dot{\phi}_2)$ is given by $V_\mathrm{gd1} = \mathcal{J} \cdot V_\mathrm{halo}$. The proper motions $\mu_{\phi_j}$ are then given by $\mu_{\phi_j} = \dot{\phi}_j / r$.\vspace{10pt}

\section{Network Architecture}\label{app:network}

\begin{figure}[b]
\centering
\includegraphics[width=\linewidth,trim={0.1cm 0.1cm 0.1cm 0.1cm},clip]{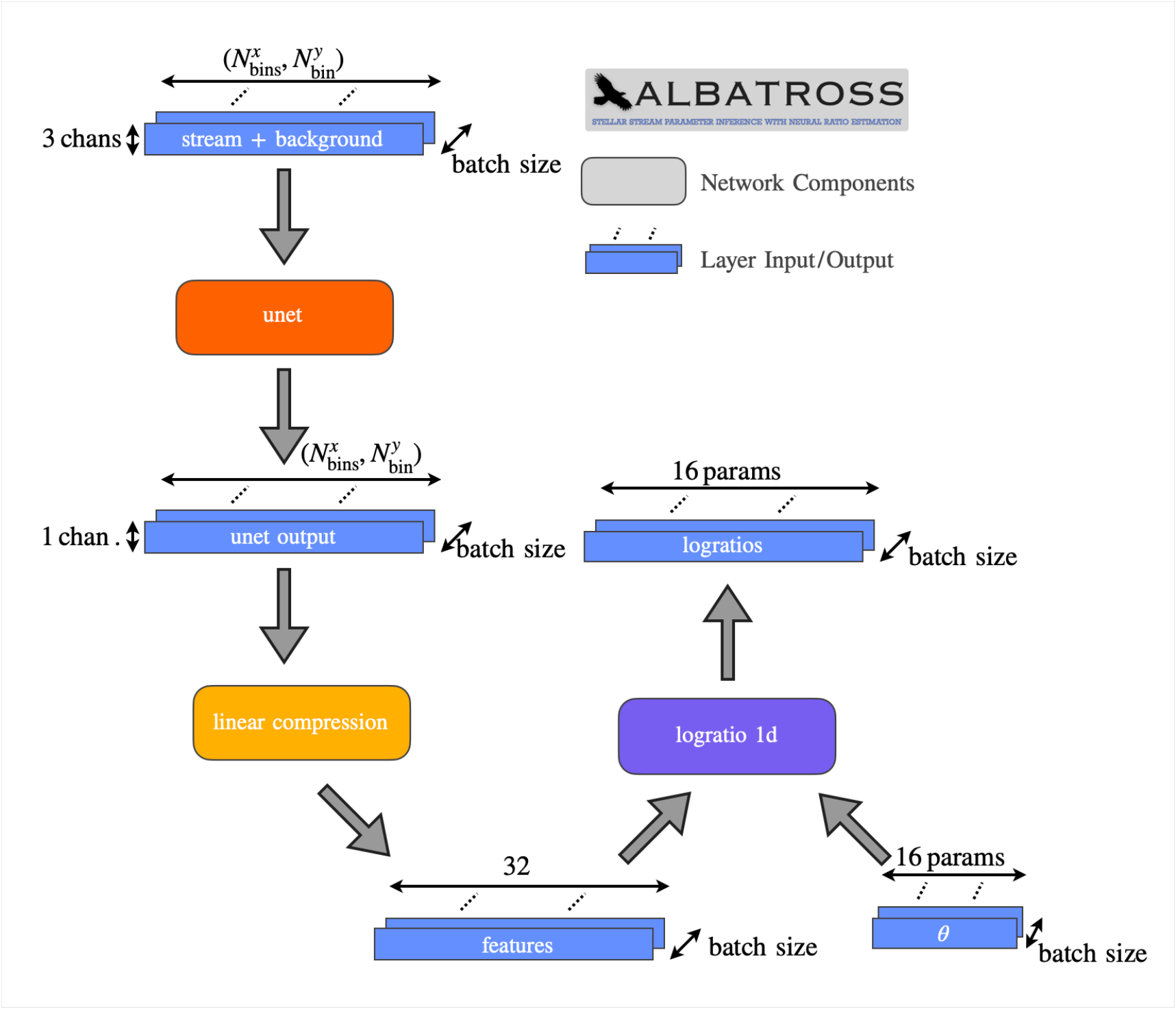}
\caption{Schematic network diagram illustrating the data processing and ratio estimation network architecture employed in \texttt{albatross}.}\label{fig:net}
\end{figure}

\vspace{-10pt}
\noindent In Fig.~\ref{fig:net}, we show the network architecture used in the \texttt{albatross} code to process output from the simulator and estimate the relevant likelihood-to-evidence ratios.

\begin{figure*}[t]
    \centering
    \includegraphics[width=\linewidth,trim={0.1cm 0.1cm 0.1cm 0.1cm},clip]{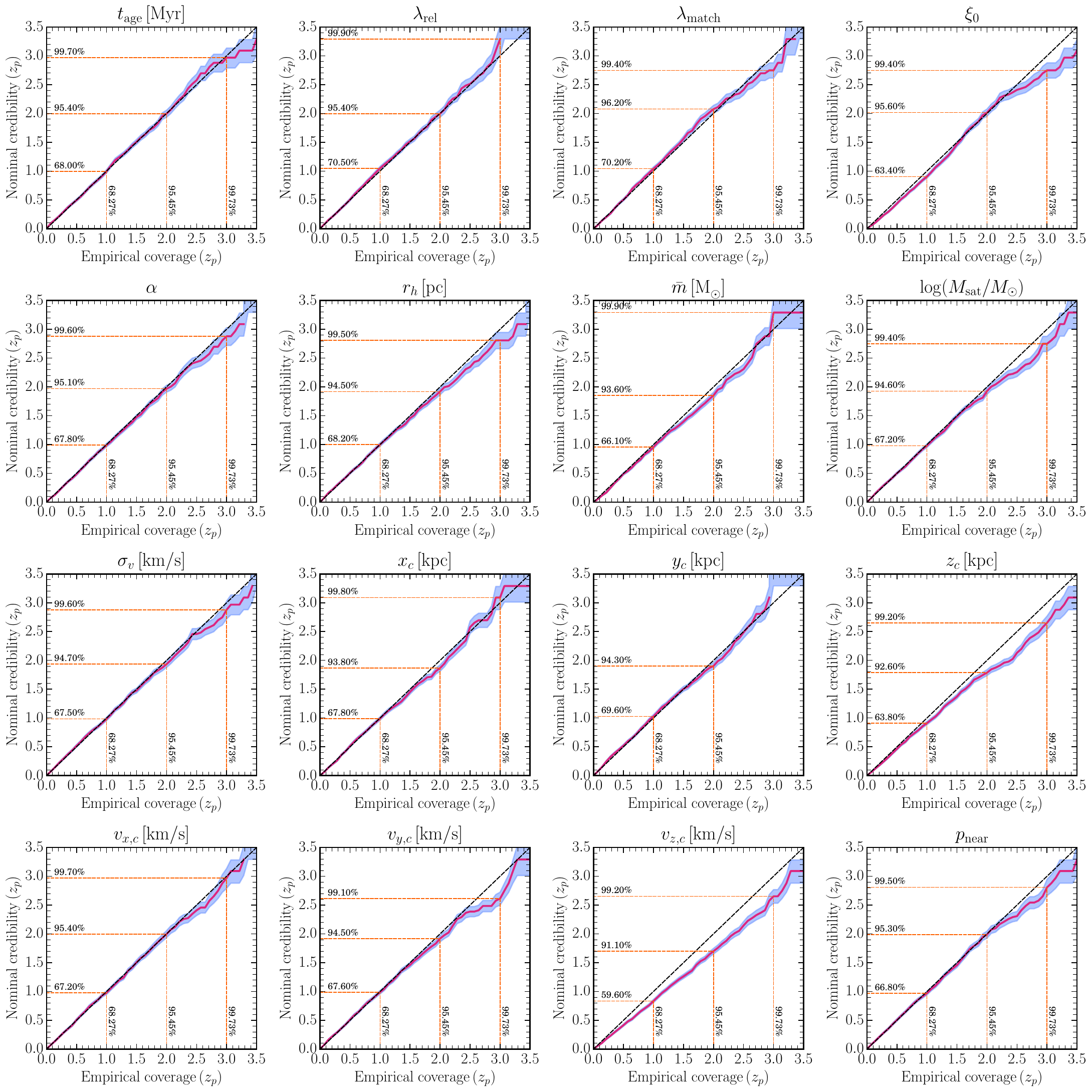}
    \caption{Coverage results for the case study given in Sec.~\ref{sec:results} for all parameters in the \texttt{sstrax} stream model. This is the same information as Fig.~\ref{fig:pp}, but with more emphasis placed on the tail regions via the definition $p = \int_{-z_p}^{z_p}\mathrm{d}z\,(1/\sqrt{2\pi}) \exp(-z^2 / 2)$. The pink curves indicate the average coverage, whilst the blue contours represent the 1$\sigma$ uncertainty of this estimate.}
    \label{fig:zz}
\end{figure*}

\begin{figure*}[t]
    \centering
    \includegraphics[width=\linewidth,trim={0.1cm 0.1cm 0.1cm 0.1cm},clip]{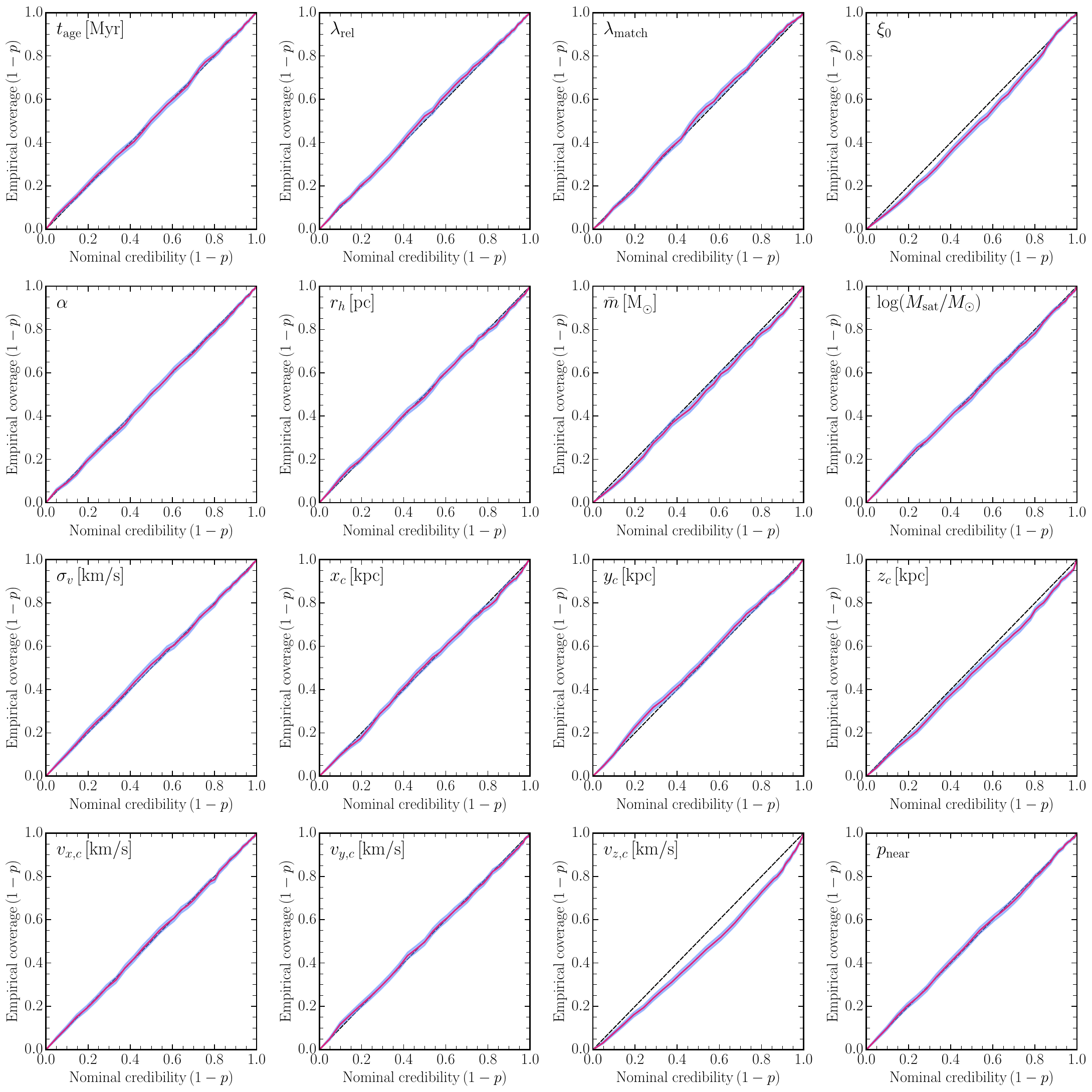}
    \caption{Coverage results for the case study given in Sec.~\ref{sec:results} for all parameters in the \texttt{sstrax} stream model. The pink curves indicate the average coverage, whilst the blue contours represent the 1$\sigma$ uncertainty of this estimate.}
    \label{fig:pp}
\end{figure*}
\end{document}